\documentclass[preprint]{aastex61}
\usepackage{amsmath}
\renewcommand\vec{\boldsymbol}

\begin{document}

\title{Modeling Emission of Heavy Energetic Neutral Atoms from the Heliosphere}
\correspondingauthor{Pawe\l{} Swaczyna}
\email{pswaczyna@cbk.waw.pl}

\author[0000-0002-9033-0809]{Pawe\l{} Swaczyna}
\affiliation{Space Research Centre of the Polish Academy of Sciences (CBK PAN), Bartycka 18A, 00-716 Warsaw, Poland}
\author{Maciej Bzowski}
\affiliation{Space Research Centre of the Polish Academy of Sciences (CBK PAN), Bartycka 18A, 00-716 Warsaw, Poland}

\begin{abstract}
Observations of energetic neutral atoms (ENAs) are a fruitful tool for remote diagnosis of the plasma in the heliosphere and its vicinity. So far, instruments detecting ENAs from the heliosphere were configured for observations of hydrogen atoms. Here, we estimate emissions of ENAs of the heavy chemical elements helium, oxygen, nitrogen, and neon. A large portion of the heliospheric ENAs is created in the inner heliosheath from neutralized interstellar pick-up ions (PUIs). We modeled this process and calculated full-sky intensities of ENAs for energies 0.2--130 keV/nuc. We found that the largest fluxes among considered species are expected for helium, smaller for oxygen and nitrogen, and smallest for neon. The obtained intensities are 50--10$^6$ times smaller than the hydrogen ENA intensities observed by \emph{IBEX}. The detection of heavy ENAs will be possible if a future ENA detector is equipped with the capability to measure the masses of observed atoms. Because of different reaction cross-sections among the different species, observations of heavy ENAs can allow for a better understanding of global structure of the heliosphere as well as the transport and energization of PUIs in the heliosphere. 
\end{abstract}


\section{Introduction}\label{sec:introduction}
Observations of Energetic Neutral Atoms (ENAs) allow for the remote diagnosis of plasma in the heliosphere \citep{gruntman_1997}. Remote sensing provides observations of various parts of the heliosphere that is evolving with the solar cycle. So far, the heliospheric ENAs have been observed by several instruments placed in outer space. Global observations of hydrogen ENA fluxes with energies  $\lesssim$6~keV are made by \emph{Interstellar Boundary Explorer} \citep[\emph{IBEX}][]{mccomas_2009a}. Observations of more energetic ENAs from the heliosphere are available from INCA on board \emph{Cassini} \citep{mitchell_1993} and from HSTOF on board \emph{SOHO} \citep{hovestadt_1995}. 

\emph{IBEX} provides full sky measurements of hydrogen ENA intensities in several energy bands and allows for the analysis of temporal variation in the signal \citep{mccomas_2009, mccomas_2012a, mccomas_2014b, mccomas_2017}. These observations showed that in addition to the globally distributed flux, an additional component, not expected prior to IBEX launch and called the \emph{IBEX} ribbon, is pronounced on an $\sim$20$\degr$ wide band that spans $\sim$300$\degr$ across the sky \citep{fuselier_2009, schwadron_2014a}. The globally distributed flux likely originates in the inner heliosheath \citep{schwadron_2011, zirnstein_2016b, zirnstein_2017}. INCA allowed for nearly full sky measurements of hydrogen ENA intensities for energies 5--55~keV \citep{krimigis_2009, dialynas_2013, dialynas_2017}, i.e., larger than \emph{IBEX} but with an inhomogeneous coverage in time. HSTOF observed hydrogen ENAs of energies 55--80 keV but only in selected directions \citep{hilchenbach_1998}.

The heliospheric ENAs are expected to be almost entirely hydrogen atoms. For this reason, the \emph{IBEX}-Hi detector was not designed to recognize chemical elements \citep{funsten_2009a}. \citet{allegrini_2008} suggested that ratios of fluxes of the ENAs with different masses could potentially be recognized on a statistical basis from the event coincidence ratios in the detector section of \emph{IBEX}-Hi. However, this method requires that fluxes of atoms of different elements are of comparable magnitudes, which is not the case for the heliosphere, and consequently, there is no confirmation of heavy ENA observations from \emph{IBEX}-Hi. Extraction of the helium ENA signal was possible from HSTOF measurements during quiet solar times \citep{czechowski_2012}. The observations were gathered in large swaths of the sky and allowed for the determination of helium ENA spectra in four ecliptical sectors for energies 28--58~keV/nuc. 

\citet{grzedzielski_2010a} modeled emission of heavy ENAs\footnote{Hereafter, ENAs of chemical elements other than hydrogen are called heavy ENAs.} from the inner heliosheath. In their model, the source population for ENAs are solar wind ions emitted from the solar corona. The authors considered various reactions that change the charge state of heavy ions of carbon, nitrogen, oxygen, magnesium, silicon, and sulfur emitted by the solar corona. This model was extended in a follow-up paper by \citet{grzedzielski_2013}, where helium ions were considered. In this paper, contribution from helium pick-up ions (PUIs) as an additional component of ions in the inner heliosheath was added. \citet{czechowski_2012} also modeled helium ENA emission and compared the result with the HSTOF observations. However, the modeled signal was significantly larger than the observed one. \citet{grzedzielski_2014} extended the model to the higher energies and were able to match the HSTOF observations using \emph{Voyagers} observations of the PUIs spectrum in the inner heliosheath \citep{stone_2008}, as well as the thickness of the inner heliosheath obtained from in situ observations \citep{burlaga_2013, stone_2013}.

The other component of the ENA emission is the \emph{IBEX} ribbon. \citet{swaczyna_2014} made assessments of helium ENA emission in two alternative mechanisms of the \emph{IBEX} ribbon. They showed that helium ENA intensities from the ribbon are typically smaller than the inner heliosheath emission in the hypothesis of the secondary ENA emission from the outer heliosheath \citep{schwadron_2009, heerikhuisen_2010}. Alternatively, they found that the extraheliospheric sources, as suggested by \citet{grzedzielski_2010} for the \emph{IBEX} ribbon, may be an important source of helium ENAs. Recent results suggest that the \emph{IBEX} ribbon is created by the secondary ENA mechanism \citep{zirnstein_2015a, zirnstein_2015, zirnstein_2016a, swaczyna_2016, swaczyna_2016b, mccomas_2017}. 

In a previous paper \citep[Paper I]{swaczyna_2017}, we modeled helium ENA emission using both components: the globally distributed flux from the inner heliosheath and the \emph{IBEX} ribbon formed in the outer heliosheath through the secondary ENA mechanism. The present paper complements the estimations of heavy ENA emission from the inner heliosheath with calculations made for three additional chemical elements: oxygen, nitrogen, and neon. These elements are chosen due to the largest abundance of the neutral atoms in the local interstellar medium \citep{frisch_2011}.

However, oppositely to the analysis presented in Paper I, the \emph{IBEX} ribbon signal from the outer heliosheath is not included. In Paper I, we found that it is generally small compared to the inner heliosheath emission, and certainly not dominant, as it is for hydrogen. For heavier elements, it is likely even less important due to the lack of these elements in the neutralized components of the supersonic solar wind \citep{grzedzielski_2010a}. Nonetheless, this mechanism may produce some ENAs from the PUIs that are neutralized in the heliosphere and that escape to the outer heliosheath, but this component is most likely small.

The model and details of calculations including discussion of differences compared with the method in Paper I are given in Section~\ref{sec:method}. The obtained intensities of heavy ENAs are shown as spectra and maps in Section~\ref{sec:results}, where also possibilities of detection with future instruments is provided. Discussion on the usability of heavy ENA observations for investigation of the heliospheric physics are presented in Section~\ref{sec:discussion}. The main findings and conclusions are summarized in Section~\ref{sec:summary}.

\section{Method}\label{sec:method}
The aim of this analysis is to provide estimations of intensities of ENAs of heavy elements, such as helium, oxygen, nitrogen, and neon. Due to various possible implementations of ENA detectors in future space experiments, the estimations are performed for all sky observations for a wide range of energies from 0.2 keV/nuc to 130 keV/nuc, or equivalently for speeds from 200 $\mathrm{km\,s^{-1}}$ to 5000~$\mathrm{km\,s^{-1}}$.

The method used in this paper follows the scheme presented in \citet{grzedzielski_2013, grzedzielski_2014}, and in Paper I, with some modifications discussed below. The modifications are employed to allow for the analysis of heavier elements with a limited knowledge of the reaction cross-sections. One of the major differences is that only single ionized ions are considered in this analysis. Consequently, the solar wind ions, which emerge from the solar corona in higher ionization states, are not included in this analysis. The ionization processes that increase the ionization state are also neglected.

Section~\ref{sec:modelIHS} presents the adopted model of the heliosphere, which describes the plasma flow in the inner heliosheath. Assumptions on the distribution functions of ions at the termination shock and their evolution in the inner heliosheath are discussed in Section~\ref{sec:PUIdistribution} and \ref{sec:PUIevol}, respectively. Finally, integration of ENA production along lines of sight is shown in Section~\ref{sec:ENAintensities}. 

\subsection{Model of the inner heliosheath structure and plasma flow}\label{sec:modelIHS}
The flow is assumed to be incompressible and irrotational. Therefore, the flow of the plasma in the inner heliosheath can be obtained from the scalar potential \citep{parker_1961, suess_1990}. The form and parameters are adopted following the potential $\Phi_1$ in Paper I. In this potential, the plasma bulk flow $\vec{u}$ in the inner heliosheath is given as a gradient of the scalar potential $\Phi$:
\begin{equation}
\vec{u}=-\frac{1}{\sqrt{n}}\vec{\nabla}\Phi,
\end{equation} 
where $n$ is the number density of protons in the inner heliosheath, and the potential has the form \citep{suess_1990}
\begin{equation}
\Phi=-\sqrt{n_\mathrm{i}}u_\mathrm{i} z \left[ \frac{1}{2}\left( \frac{r_\mathrm{s}}{r}\right)^3+1\right]+\sqrt{n_\mathrm{s}}u_\mathrm{s}\frac{r_\mathrm{s}^2}{r},
\end{equation}
where $n_\mathrm{i}$ is the number density of protons in the interstellar medium, $n_\mathrm{s}$ is the number density of protons in the inner heliosheath, $u_\mathrm{i}$ is the plasma speed in the local interstellar medium relative to the Sun, $u_\mathrm{s}$ is the radial component of the plasma speed downstream of the termination shock in the solar frame, which is located at the radial distance $r_\mathrm{s}$. These parameters are the same as those used previously in Paper I: $n_\mathrm{i}=0.0659$~$\mathrm{cm^{-3}}$, $n_\mathrm{s}=0.002$~$\mathrm{cm^{-3}}$, $u_\mathrm{i}=-25.8$~$\mathrm{km\,s^{-1}}$, $u_\mathrm{s}=150$~$\mathrm{km\,s^{-1}}$, and $r_\mathrm{s}=90$~au. These values allow for fair reproduction of the mean flow observed by \emph{Voyagers} \citep{richardson_2011a}, as well as the position of the heliopause along the \emph{Voyager 1} trajectory \citep{gurnett_2013}. 

This flow assumes conservation of mass along the streamlines. In particular, this means that there should not be significant mass loading to the plasma in the inner heliosheath. This assumption was recently challenged by \citet{fahr_2015}, who suggested that electrons in the inner heliosheath can be significantly hotter than previously thought, with an upper limit of $\sim$10~eV \citep{richardson_2009}. In fact, mass loading due to electron impact ionization could be large in this situation \citep{gruntman_2015}. However, the actual value of the temperature is not known and thus it is not possible to unambiguously resolve this problem.

PUIs in the heliosphere are neutralized mostly due to charge exchange with the interstellar neutrals. The most abundant of them are hydrogen and helium atoms. The density of neutral atoms is assumed constant in the inner heliosheath and equal $n_\mathrm{H}=0.1$~$\mathrm{cm^{-3}}$ and $n_\mathrm{He}=0.015$~$\mathrm{cm^{-3}}$ for hydrogen and helium, respectively \citep{witte_2004, bzowski_2009, zank_2013}.

\subsection{PUIs at the termination shock}\label{sec:PUIdistribution}
PUIs are created from the interstellar neutral atoms in the inner heliosphere mostly from two possible processes:
\begin{align*}
\mathrm{Z^0\,+\,}h\nu & \rightarrow \mathrm{ Z^+ }\\
\mathrm{Z^0\,+\,H^+} & \rightarrow \mathrm{Z^+ + H^0}
\end{align*}
where Z$^0$ represents an interstellar neutral atom and Z$^+$ is the created PUI (here Z = He, O, N, or Ne). The first of these processes is photoionization, the second is the charge exchange with the solar wind protons. Ionization by impact of hot electrons can be important in the supersonic solar wind for some species, but only close to the Sun \citep{rucinski_1991, bzowski_2013b, bzowski_2013c}. Consequently, the contribution of the PUIs created by electron ionization is negligible at the termination shock. Other charge exchange processes are less important. 

The relevance of the two reactions considered in this analysis varies for the analyzed species. Rates of these ionization processes change with time, and for the charge exchange also with heliographic latitude \citep{sokol_2016}. This study is limited to the average solar wind condition. Consequently, the solar wind parameters are averaged over the period of solar cycle 23, as derived in Paper I, based on the model from \citet{sokol_2012, sokol_2015c}. The solar wind speed is significantly larger than the bulk speed of interstellar neutrals and the thermal speeds of the solar wind protons and interstellar neutrals. Hence, the charge-exchange rate at 1 au can be calculated as
\begin{equation}
\nu_\mathrm{cx}(\theta)=\sigma_\mathrm{cx}(v_\mathrm{SW,0}(\theta))v_\mathrm{SW,0}(\theta)n_\mathrm{SW,0}(\theta),
\end{equation}
where $\sigma_\mathrm{cx}$ is the charge-exchange cross-section (see Appendix~\ref{app:cxnp}), $v_\mathrm{SW,0}(\theta)$ is the solar wind speed, and $n_\mathrm{SW,0}(\theta)$ is the solar wind density at 1~au for heliographic latitude $\theta$. 

The ionization rates of the considered species for average solar wind conditions are presented in Table~\ref{tab:ionrates}. The charge-exchange rates ($\nu_\mathrm{cx}(\theta)$) are provided for heliographic latitudes 0$\degr$, 30$\degr$, 60$\degr$, and 90$\degr$. The table also provides the average charge-exchange rate ($\bar{\nu}_\mathrm{cx}$), and the photoionization rates ($\nu_\mathrm{ph}$). The photoionization rates are calculated as an average over solar cycle 23 based on photoionization rates from \citet{sokol_2016}, except for nitrogen, for which the value provided by \citet{rucinski_1996} is used. 

\begin{deluxetable*}{lrrrrrrrr}
	\tablecaption{Charge-exchange and photoionization rates at 1 au for selected heliographic latitudes ($10^{-7}$~s$^{-1}$)\label{tab:ionrates}}
	\tablehead{
		\colhead{} & 
		\colhead{$\nu_\mathrm{cx}(0\degr)$} & 
		\colhead{$\nu_\mathrm{cx}(30\degr)$} & 
		\colhead{$\nu_\mathrm{cx}(60\degr)$} & 
		\colhead{$\nu_\mathrm{cx}(90\degr)$} & 
		\colhead{$\bar{\nu}_\mathrm{cx}$} &
		\colhead{$\nu_\mathrm{ph}$} &
		\colhead{$\bar{\nu}_\mathrm{cx}+\nu_\mathrm{ph}$} 
	}
	\startdata
	\text{He} & 0.002 & 0.005 & 0.011 & 0.016 & 0.006 & 1.034 & 1.040 \\
	\text{O}  & 2.355 & 2.057 & 1.767 & 1.699 & 2.044 & 3.921 & 5.965 \\
	\text{N}  & 4.517 & 3.740 & 3.021 & 2.821 & 3.716 & 4.380 & 8.096 \\
	\text{Ne} & 0.022 & 0.059 & 0.139 & 0.195 & 0.074 & 3.079 & 3.153 \\
	\enddata
\end{deluxetable*}

The density distributions of the interstellar neutrals in the supersonic solar wind are not homogeneous due to ionization losses and gravitational deflection of atom trajectories. Consequently, the density of interstellar neutrals depends on time and position in the heliosphere. The potential symmetries of the distributions are lost due to the dependence of ionization rates on heliographic latitude. However, for the heavy elements that are considered in this study, the cold model of interstellar neutrals \citep[e.g.][]{blum_1970} can be employed. 

\citet{fahr_1971} showed that the cold model reproduces the densities quite well except for a cone around the downwind direction. A half angle of this cone is given by \citep{thomas_1978}
\begin{equation}
\gamma=\arctan \frac{\sqrt{2kT_\mathrm{i}/m}}{u_i},
\end{equation}
where $k$ is the Boltzmann constant, $T_\mathrm{i}$ is the temperature of the interstellar neutrals, and $m$ is the atomic mass. For the temperature $T_\mathrm{i}=7440$~K \citep{bzowski_2015d}, the resulting angle is equal to $\sim$12$\degr$ for helium and 5.5$\degr$--6.5$\degr$ for the other considered species. \citet{feldman_1972} derived an analytic formula for the density in the downwind direction for the limit of small but not zero temperature, as assumed for the cold model. Here, the density of the interstellar neutrals is calculated based on the cold model, except for the cone defined by the cone-angle $\gamma$. Within this cone the interpolation between the cold model and the result of the formula given by \citet{feldman_1972} is used. 

The number densities of interstellar neutrals are taken from the ionization model given by \citet{frisch_2011} and as model 26 in \citet{slavin_2008}. These densities correspond to the medium far away from the heliosphere and the gas can be substantially filtered in the outer heliosheath. This issue was discussed for heavy interstellar atoms by \citet{cummings_2002}, \citet{izmodenov_2004}, and \citet{muller_2004}. The filtration is likely the largest for oxygen among the considered species. However, these authors obtained substantially different filtration factors, ranging from $\sim$0.7 to $\sim$1.3. Consequently, the filtration effect is neglected in this analysis, and the interstellar values are directly applied to the cold model. Finally, the densities for the cold model far away from the Sun are adopted as follows: helium -- $1.5\times 10^{-3}$~$\mathrm{cm^{-3}}$, oxygen -- $7.3\times 10^{-5}$~$\mathrm{cm^{-3}}$, nitrogen -- $9.2\times 10^{-6}$~$\mathrm{cm^{-3}}$, neon -- $6.6\times 10^{-6}$~$\mathrm{cm^{-3}}$. The other heavy atoms are much less abundant ($<2\times10^{-7}$~$\mathrm{cm^{-3}}$) in the local interstellar medium and thus they are not considered in this analysis.

In the cold model, the ionization rates are assumed constant in time and isotropic, so they depend only on the distance $r$ from the Sun $r$ as $\nu(r)=\nu_0 (r_0/r)^2$, where $r_0=1$~au, and $\nu_0$ is the ionization rate at 1~au. The sum of the averaged charge-exchange rate and photoionization rate is used for the considered species. 

The flux of PUIs at the termination shock in a selected direction is calculated from the formula
\begin{equation}
S_\mathrm{PUI}(\psi,\theta)=\frac{1}{r_\mathrm{s}^2}\int_0^{r_\mathrm{s}} n_\mathrm{ISN}(r,\psi)\left[ \nu_\mathrm{cx}(r,\theta)+\nu_\mathrm{ph}(r)\right]r^2dr,
\end{equation}
where $\psi$ is the angular distance from the inflow direction, $\theta$ is the heliographic latitude, $n_\mathrm{ISN}$ is the density of the interstellar neutrals obtained from the model, and $r_\mathrm{s}$ is the distance to the termination shock from the Sun. From continuity, the density of PUIs downstream of the shock can be calculated as $n_\mathrm{PUI}=S_\mathrm{PUI}/u_\mathrm{s}$. In this analysis, only the case of spherical termination shock is considered (for a discussion of this issue see Paper I).

The acceleration of PUIs at the termination shock is a source of high energy tail of their distribution functions \citep{zank_1996}. It is useful to split PUIs into two populations: the transmitted PUIs and the reflected PUIs. We construct the distribution function in the same way as presented in Paper I. Namely, for the transmitted PUIs, it is assumed that they follow the Maxwell--Boltzmann distribution with the mean energy corresponding to the difference between the kinetic energies resulting from the bulk speeds upstream and downstream of the termination shock. The reflected PUIs are assumed to follow the $\kappa$-distribution with $\kappa=1.65$ \citep{decker_2005}. 

In Paper I, the mean energy of the reflected helium PUIs was determined from fitting to the \emph{Voyager 1} measurements at $\sim$1 MeV/nuc from the first two months of 2007 presented by \citet{stone_2008}. The obtained mean energy was 19.1~keV/nuc, assuming that 5\% of the total PUI population is reflected. \citet{stone_2008} also showed results for oxygen, and the same procedure is applicable. For the same part of the reflected PUIs (5\%), the resulting energy is 19.7~keV/nuc. This suggests that the process of the acceleration at the termination shock is similar for different species. Hence, for all of the analyzed species, the distribution function for the reflected PUIs is assumed the same with the mean energy equal 19.4~keV/nuc. Finally, the distribution function in the velocity space of PUIs downstream of the termination shock has the form
\begin{equation}
f_\mathrm{PUI,TS}(v)=n_\mathrm{PUI}\left[(1-\eta_\mathrm{r})f_\mathrm{MB}(v_\mathrm{t},v)+\eta_\mathrm{r}f_\kappa(v_\mathrm{r},v) \right],
\end{equation}
where $\eta_\mathrm{r}=0.05$ is portion of the reflected PUIs, $f_\mathrm{MB}(v_\mathrm{t},v)$ is the Maxwell--Boltzmann distribution with thermal speed $v_\mathrm{t}$, and $f_\kappa(v_\mathrm{r},v)$ is the $\kappa$-distribution with $\kappa=1.65$ and $v_\mathrm{r}$ is the characteristic speed for the reflected population. The relation of the thermal and characteristic speeds to the respective energies and the definition of the distribution are the same as provided in Paper I.

The mean energies of the transmitted PUIs, obtained as described above, range from 0.55 keV/nuc at the solar equator to 1.35 keV/nuc close to the Sun's poles. For the reflected PUIs, the energies are scaled using those for the transmitted PUIs, so at the heliographic latitude of \emph{Voyager 1} the mean energy is 19.4~keV/nuc. In consequence, the mean energies of reflected PUIs range from 12 keV/nuc to 32 keV/nuc. 

\subsection{PUIs in the inner heliosheath}\label{sec:PUIevol}
Assuming that the velocity diffusion and spatial diffusion are inefficient, the transport equation in the inner heliosheath within the considered incompressible plasma model reduces to (see Paper I)
\begin{equation}
\frac{\partial f_\mathrm{PUI}}{\partial t}=\left.\frac{\delta f_\mathrm{PUI}}{\delta t}\right|_\mathrm{c}. \label{eq:transport}
\end{equation}
The right-hand side of the equation denotes the collision term. This study considers only the neutralization of PUIs on the interstellar neutrals. For helium PUIs, neutralization on hydrogen and helium atoms is included, for the other elements neutralization is only on hydrogen atoms. Hence, the collision terms are
\begin{equation}
\left.\frac{\delta f_\mathrm{PUI}}{\delta t}\right|_\mathrm{c}=-\left(\sigma_\mathrm{cx,H} v_\mathrm{rel} n_\mathrm{H} +\sigma_\mathrm{cx,He} v_\mathrm{rel} n_\mathrm{He}\right) f_\mathrm{PUI}, \label{eq:coll}
\end{equation}
where $\sigma_\mathrm{cx,H}$ ($\sigma_\mathrm{cx,He}$) denotes the charge-exchange cross-sections with neutral hydrogen (helium) atoms, $v_\mathrm{rel}$ is the relative speed between ions and neutrals, and $n_\mathrm{H}$ ($n_\mathrm{He}$) is the number density of neutral hydrogen (helium) atoms. The charge exchange with neutral helium atoms has a significant contribution for neutralization of helium PUIs due to its resonant character, especially for lower energies $\lesssim$5 keV/nuc. For other species, the contribution from this reaction can be neglected because the density of neutral helium is one order of magnitude smaller than that of neutral hydrogen. The cross-sections used for the charge-exchange processes with hydrogen are presented in Appendix~\ref{app:cxih}. For the charge exchange between the helium ion and the helium atom, we use the cross-section from \citet{barnett_1990}. 

As in Paper I, also the contribution of PUIs created in the inner heliosheath is neglected here. These PUIs have relatively small velocities resulting from the relative motion of the interstellar neutrals and the plasma flow. The typical relative speed is of the order of $\sim$150~$\mathrm{km\, s^{-1}}$, so freshly ionized PUIs have energies below the adopted lower limit of speeds considered in this analysis. Moreover, due to the lack of efficient acceleration processes in the inner heliosheath they remain below this limit throughout the inner heliosheath in our calculations.

Equation~\eqref{eq:transport} with the collision term given by Equation~\eqref{eq:coll} can be easily solved:
\begin{equation}
f_\mathrm{PUI}(v,t)=\exp\left[ -\int_0^t \left( \sigma_\mathrm{cx,H} v_\mathrm{rel} n_\mathrm{H} +\sigma_\mathrm{cx,He} v_\mathrm{rel} n_\mathrm{He}\right)dt'\right]f_\mathrm{PUI,TS}(v), \label{eq:pui_evo}
\end{equation}
where $t$ is the time of plasma advection from the termination shock. This equation describes evolution of the distribution function of PUIs along the flow lines. In the subsequent section, the distribution function of PUIs ($f_\mathrm{PUI}(v,\vec{r})$) is considered to be a function of position in space ($\vec{r}$). This means that it is the distribution function at a connected point at the termination shock, evolved along the flow line as given by this equation. The integrand is position dependent due to the variation of relative velocities along the flow line.

Here, the evolution of the distribution function is calculated along the flow lines as a function of energy. Consequently, the distribution functions for a given flow line do not need to follow any defined distribution further in the inner heliosheath. However, we consider them to be isotropic in the velocity space. For considered energy $E$, which corresponds to speed $v=\sqrt{2E/m_\mathrm{p}}$, $f_\mathrm{PUI}(v,t)$ describes a spherical shell in the velocity space. Consequently, the mean relative speed averaged over all possible orientations of the velocity is
\begin{equation}
v_\mathrm{rel}=\frac{1}{4\pi}\int \big|\vec{v}+\vec{u}-\vec{u}_\mathrm{i}\big|d \Omega
=\frac{ v^2 + (\vec{u}-\vec{u}_\mathrm{i})^2 + 2\max(v,|\vec{u}-\vec{u}_\mathrm{i}|)^2}{3\max(v,|\vec{u}-\vec{u}_\mathrm{i}|)},
\end{equation}
where $v$ is the speed in the plasma frame, $\vec{u}$ is the plasma bulk velocity in the Sun's frame, and $\vec{u}_\mathrm{i}$ is the interstellar neutral inflow velocity. The function $\max(\cdot,\cdot)$ returns the greater of its arguments. The temperature of interstellar neutrals $\sim$7500~K \citep{bzowski_2015d, mccomas_2015c} corresponds to the characteristic velocities of 6.8~$\mathrm{km\,s^{-1}}$ and 3.4~$\mathrm{km\,s^{-1}}$ for hydrogen and helium, respectively. Therefore, the thermal speed of interstellar neutrals is neglected because it is small compared to the other speeds in the formula, and thus it is a cold ($T=0$~K) gas approximation. More generally, the formula given by \citep{ripken_1983} should be used to take the thermal component into account. The plasma bulk velocity varies along the flow line. 
Therefore, the relative velocity in the integrand of Equation~\eqref{eq:pui_evo} is a function of time, and thus the integral needs to be calculated numerically. Fortunately, based on several numerical trials, we found that eventually the calculated intensity of ENAs is accurate within $\pm$3\% if the mean relative velocity is adopted as the average along the flow line. Consequently, the integrand does not need to be evaluated at multiple points along the flow line and Equation~\eqref{eq:pui_evo} takes the form
\begin{equation}
f_\mathrm{PUI}(v,t)=e^{-\left(\sigma_\mathrm{cx,H} v_\mathrm{rel} n_\mathrm{H} +\sigma_\mathrm{cx,He} v_\mathrm{rel} n_\mathrm{He}\right)t}f_\mathrm{PUI,TS}(v), \label{eq:pui_evo2}
\end{equation}
where $v_\mathrm{rel}$ is the relative speed calculated for the averaged differences of the bulk velocities. This approximation is one of the main differences compared with the study in Paper I.

\subsection{ENA intensities at 1 au}\label{sec:ENAintensities}
The charge exchange of PUIs with interstellar neutrals is a source for ENAs. The ENA intensities from direction $\vec{\Omega}$ can be calculated from the integral in the velocity space:
\begin{equation}
 j_\mathrm{ENA}^{\mathrm{IHS},v}(\vec{\Omega},v)=\int_{r_\mathrm{s}}^{r_\mathrm{hp}}w_\mathrm{sur}(r,v)v^2f_\mathrm{PUI}\left(v',r\vec{\Omega}\right)\left(\sigma_\mathrm{cx,H} v_\mathrm{rel} n_\mathrm{H} +\sigma_\mathrm{cx,He} v_\mathrm{rel} n_\mathrm{He}\right) dr, \label{eq:jena}
\end{equation}
where $r_\mathrm{hp}$ is the distance to the heliopause, $v$ is the ENA speed in the Sun's frame, $v'$ is the PUI speed in the plasma frame, $w_\mathrm{sur}(r,v)$ is the survival probability for an atom traveling from distance $r$ to reach the observer at $r_0=1$~au. Note that only for helium, the charge exchange with the interstellar helium is included in the calculations.

The reactions that ionize the ENAs are the same reactions as those discussed for ionization of the interstellar neutrals in the supersonic solar wind. The survival probability can be calculated as
\begin{equation}
	w_\mathrm{sur}(r,v)=\exp\left[-\frac{\nu_\mathrm{ph}r_0(1-r_0/r)+\int_{r_0}^{r}\sigma_\mathrm{cx}v_\mathrm{rel}n_\mathrm{p}dr}{v}\right], \label{eq:survival}
\end{equation}
where $\nu_\mathrm{ph}$ is the photoionization rate  at 1 au. For the charge exchange, the same reactions are used (see Appendix~\ref{app:cxih}), but now the relative speed is calculated including the ENA velocities. In the inner heliosheath, the relative speed is adopted as an average of $v_\mathrm{rel}=\left|\vec{v}-\vec{u} \right|$ along the ENA path from the place of origin to the termination shock, and the proton density is constant at $n_\mathrm{p}=0.002$~$\mathrm{cm^{-3}}$. In the supersonic solar wind, the equivalent expressions are the following: $v_\mathrm{rel}=v+v_\mathrm{SW,0}$ and $n_\mathrm{p}=n_\mathrm{SW,0}(r_0/r)^2$, where $v_\mathrm{SW,0}$ and $n_\mathrm{SW,0}$ are the solar wind bulk speed and density at 1~au. With these assumptions, the integral in Equation~\eqref{eq:survival} can be split into two parts that are integrated analytically.

The integral given by Equation~\eqref{eq:jena} is the only one that needs to be calculated numerically. A reasonable accuracy is obtained using an equally spaced grid with $\Delta r=2~$~au. However, if the distance between the heliopause and the thermination shock is larger than 100~au, then up to 50~au from the termination shock the grid is created as described above, and beyond this the spacing increases by 2\% with each subsequent grid point. This procedure is used to limit the number of grid points in the downwind direction. Nonetheless, this procedure allows us to obtain an accuracy of $\sim$1\%. This value is estimated based on results for a much denser grid for several directions. 

\section{Results}\label{sec:results}
\begin{figure}
  \epsscale{.52}
  \plotone{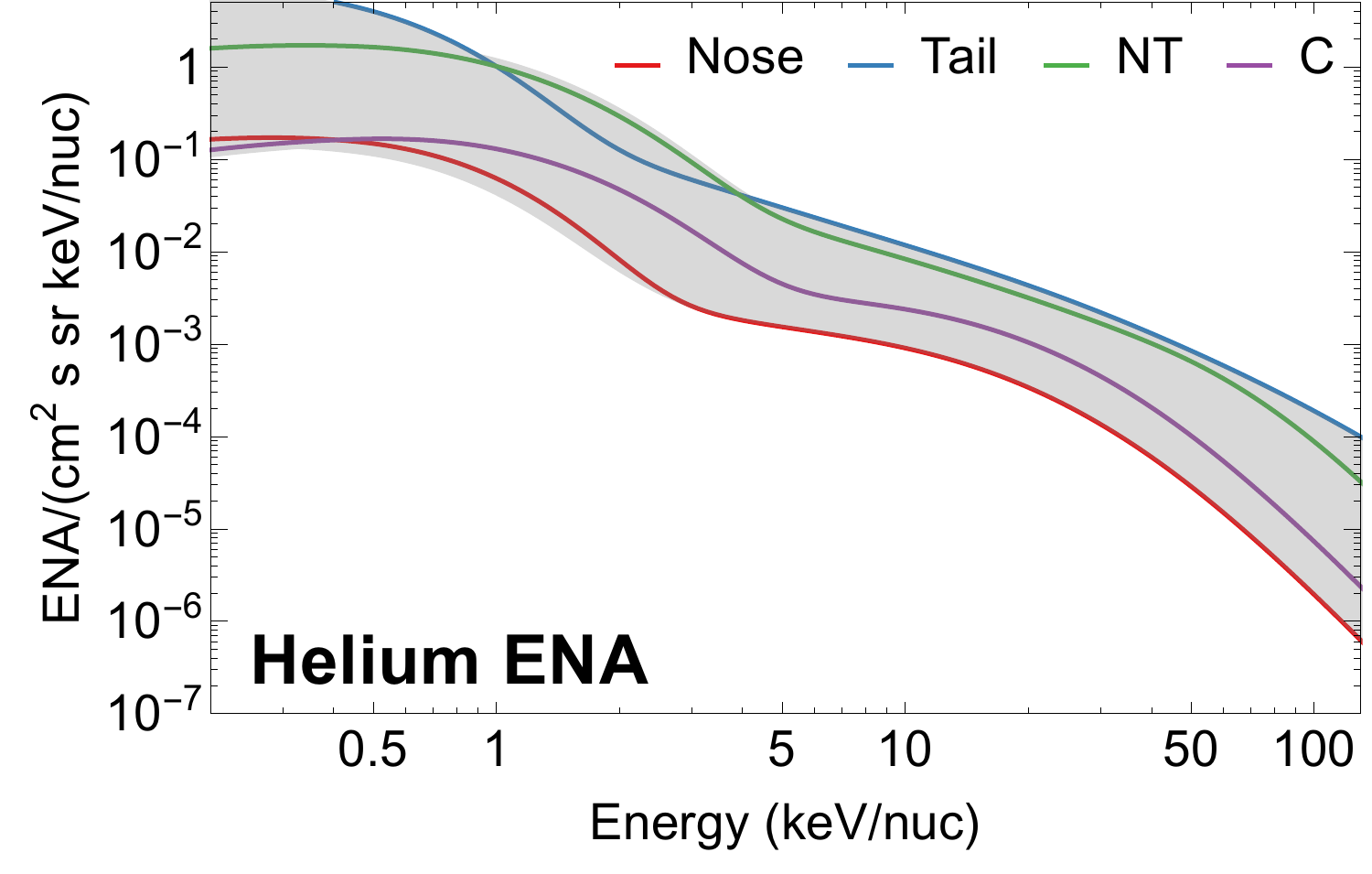}
  
  \plotone{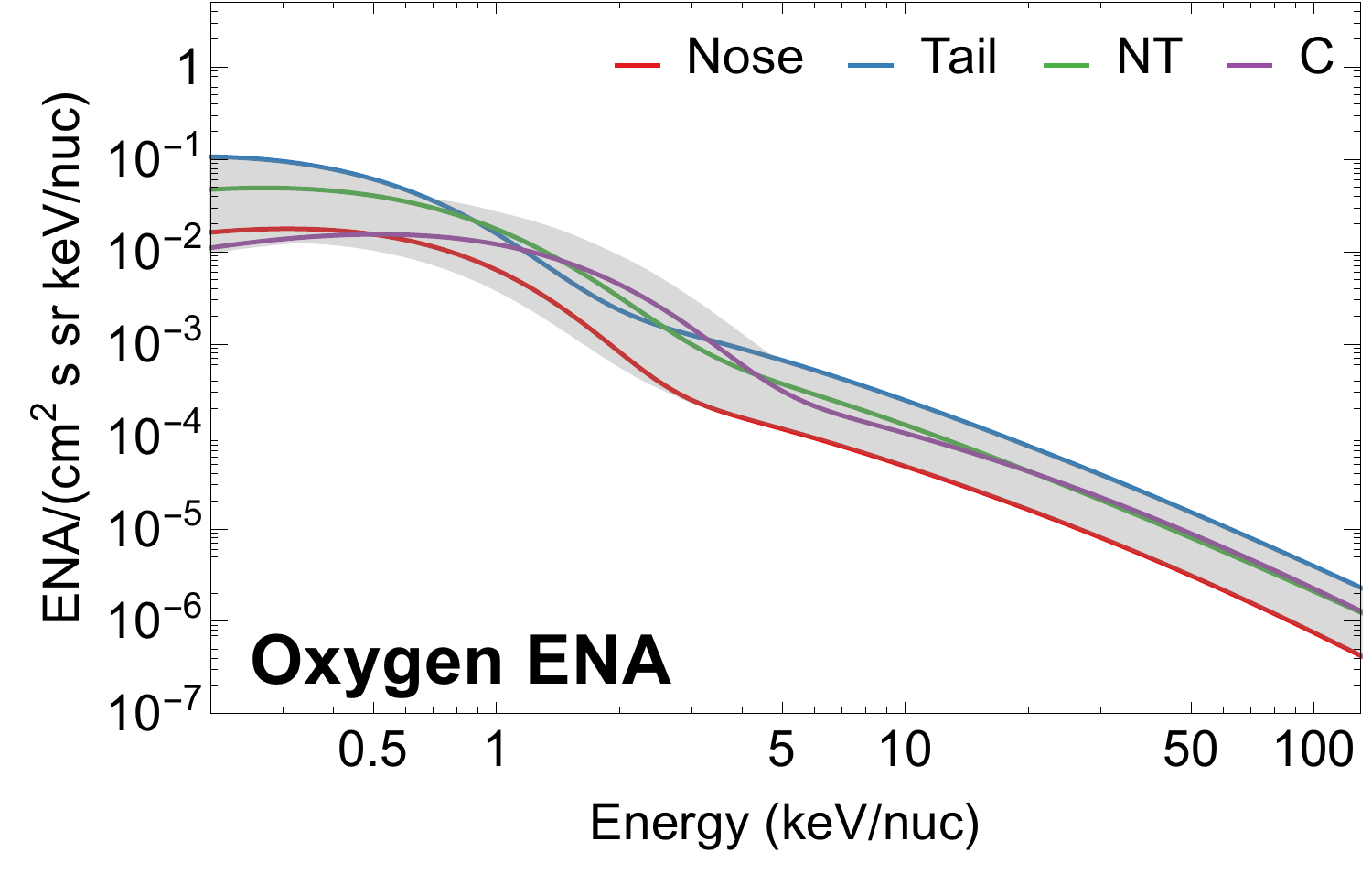}
  
  \plotone{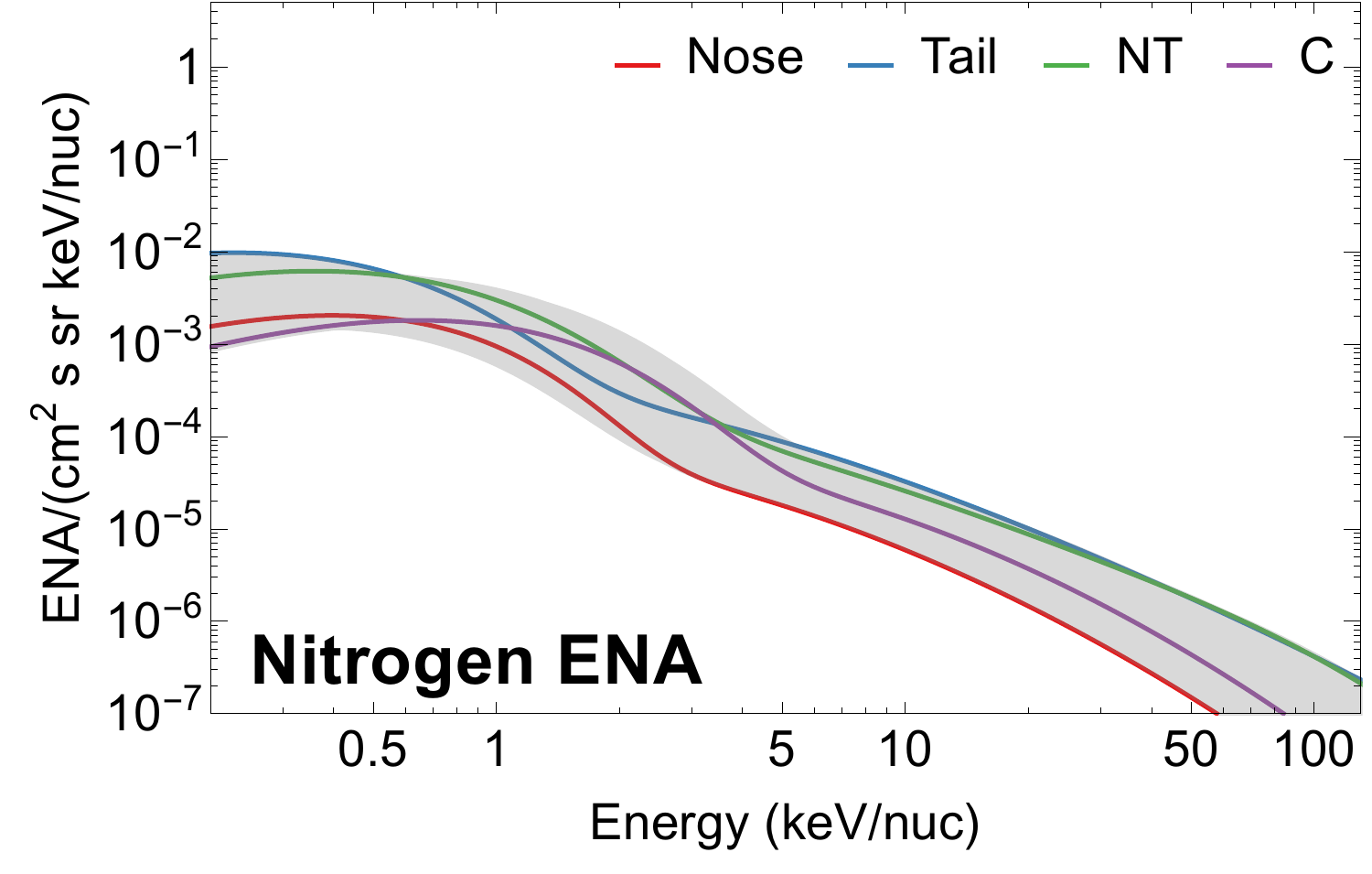}
  
  \plotone{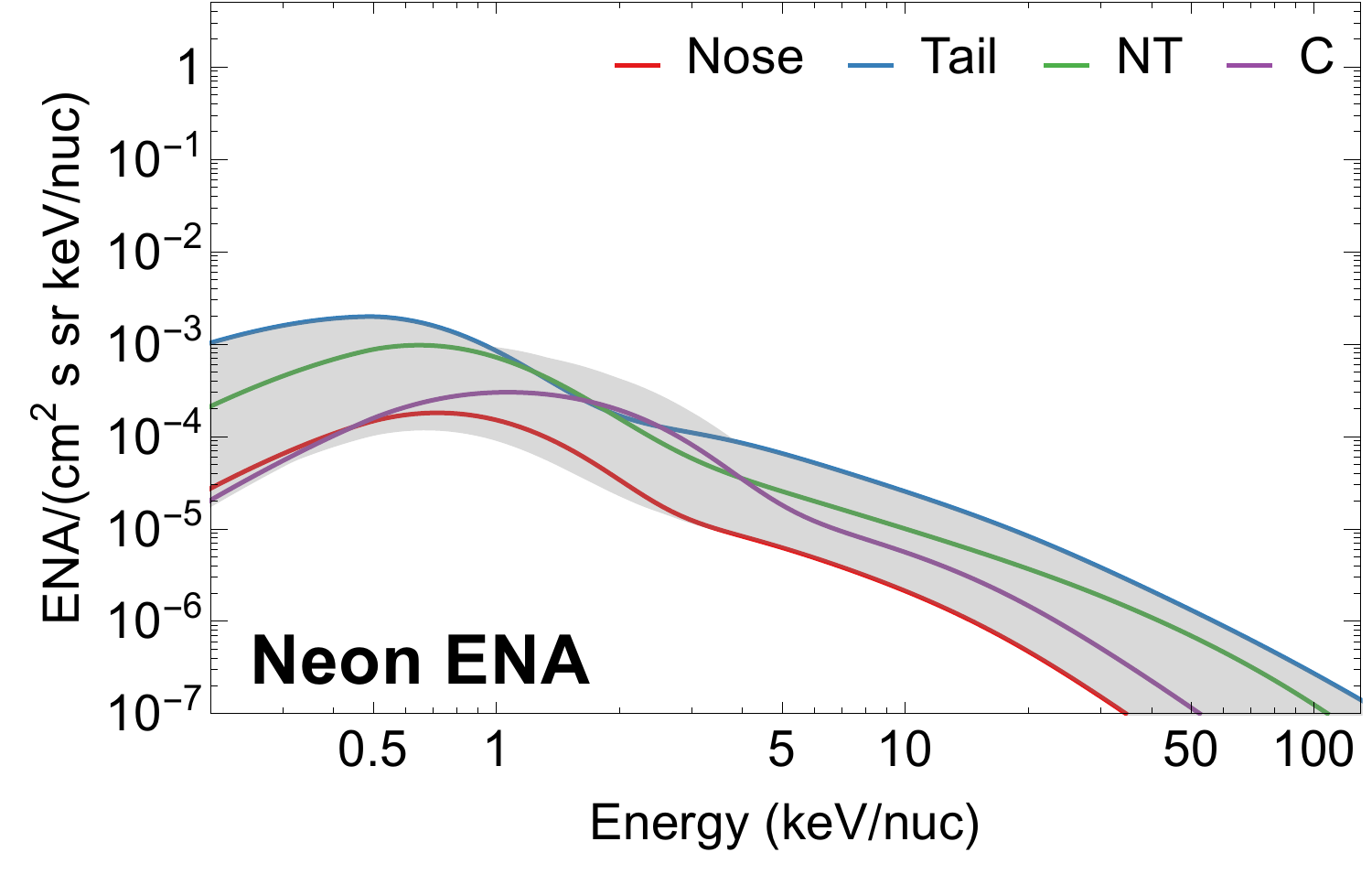}
  
  \caption{Intensity spectra of ENAs of heavy chemical elements (top to bottom): helium, oxygen, nitrogen, and neon. The gray band represents the range of the intensities for all directions in the sky. The color lines are the spectra in some selected directions (see the text). \label{fig:spectra}}
 \end{figure}

The ENA intensities are derived for four chemical elements: helium, oxygen, nitrogen, and neon, for which the expected contribution from the interstellar PUIs is the largest. The spectra of the intensities are calculated for a wide energy range from 0.2 keV/nuc to 130 keV/nuc. In Figure~\ref{fig:spectra}, intensity spectra for the analyzed elements are presented. The gray bands show the range of the spectra from all directions in the sky. Additionally, selected directions are featured with separate color lines, as the upwind (Nose) and downwind (Tail) directions. Additionally, the direction shifted toward the North Sun's Pole from the downwind direction (NT) by 10$\degr$ is also shown. This was the direction selected in the analysis of helium ENAs as a direction in which the flux is the largest for moderate energies (Paper I). The last selected direction C is one of the crosswind directions (i.e., directed 90$\degr$ away from the downwind direction) that is located at heliographic latitude 60$\degr$. All of these points are also plotted at the maps of ENA emission presented in the subsequent figures. 
 
The widest range of the intensity spectra is expected for helium ENAs. For energies $\sim$3 keV/nuc, the ratio of the intensities in the directions of the highest and the lowest intensities can reach a factor of $\sim$50. As explained in Paper I, this is a result of the long mean free paths of helium ions in the inner heliosheath. For the other considered species, the ratio is generally smaller, but still larger than that obtained in hydrogen observations made by \emph{IBEX} \citep{mccomas_2017}. 

\begin{figure*}
  	\epsscale{0.96}
  	\plottwo{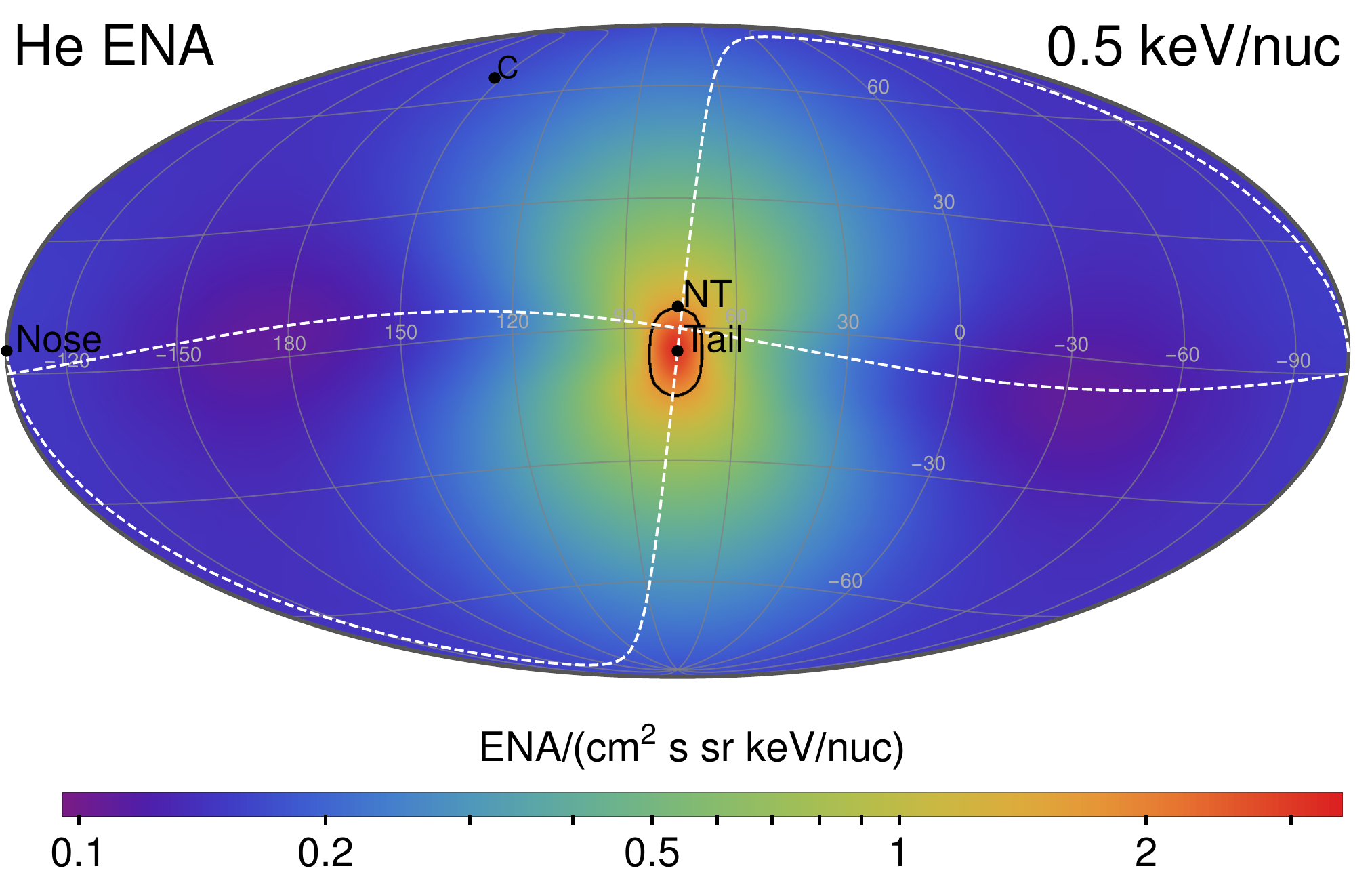}{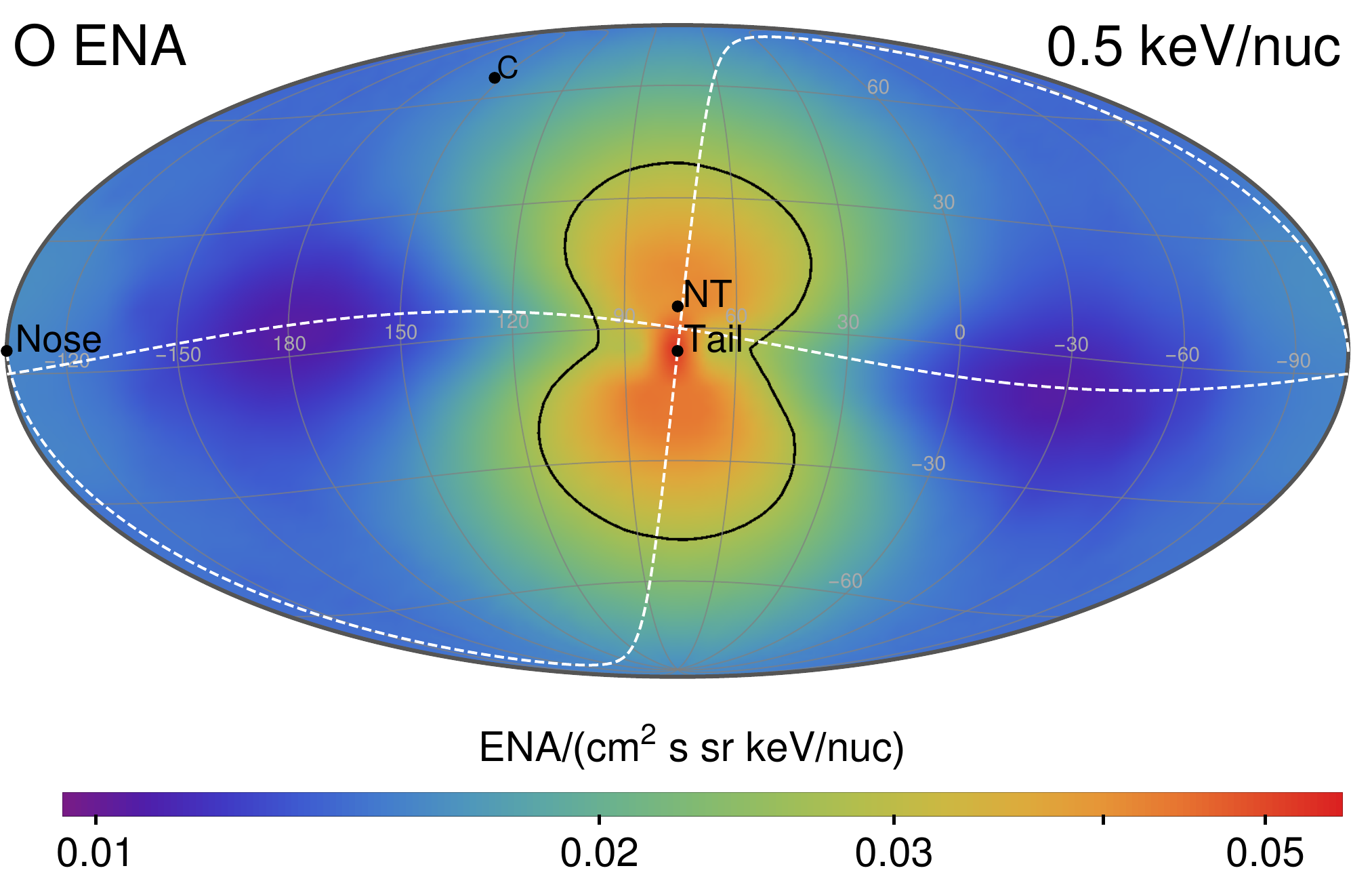}
  	\plottwo{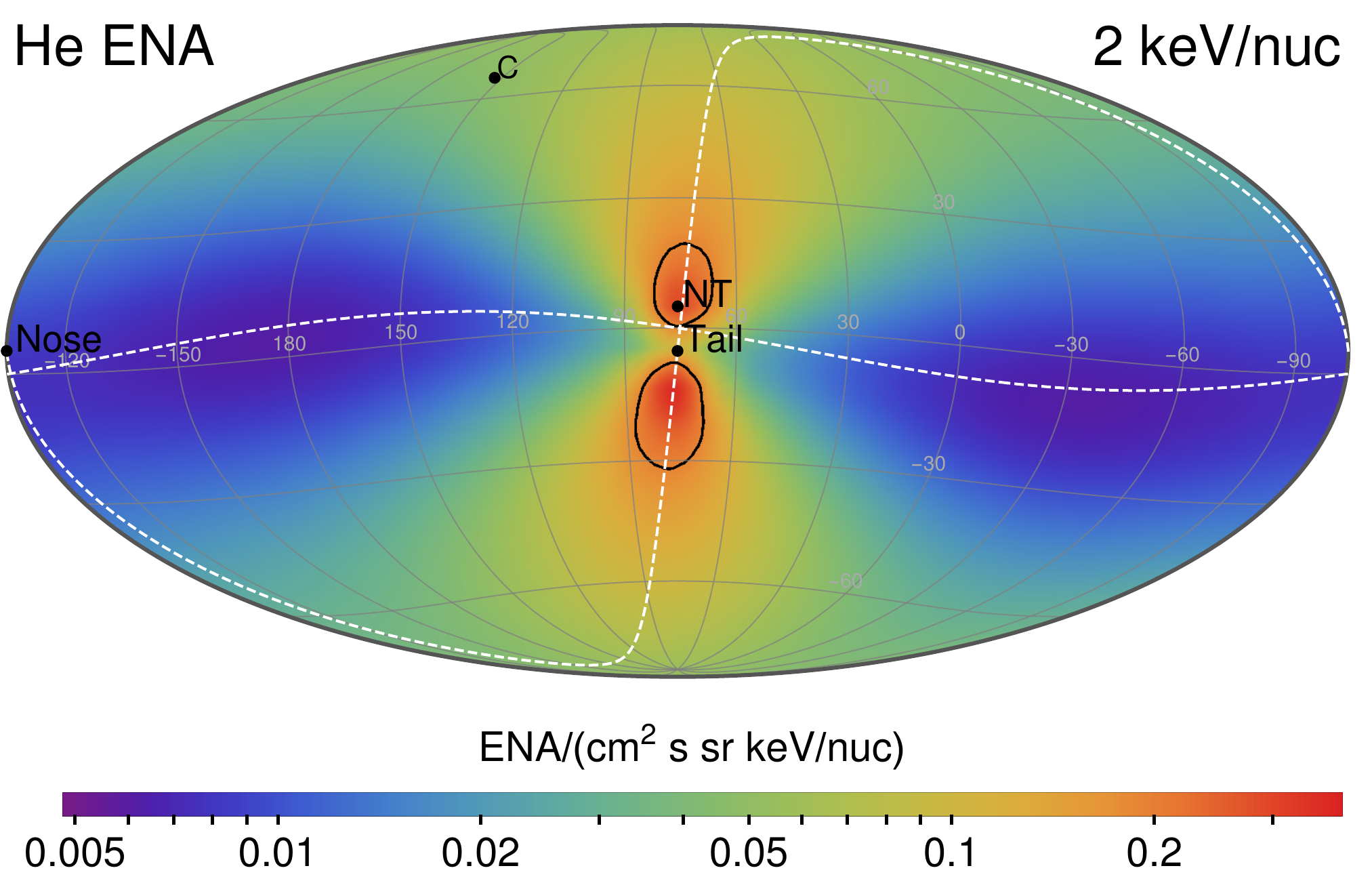}{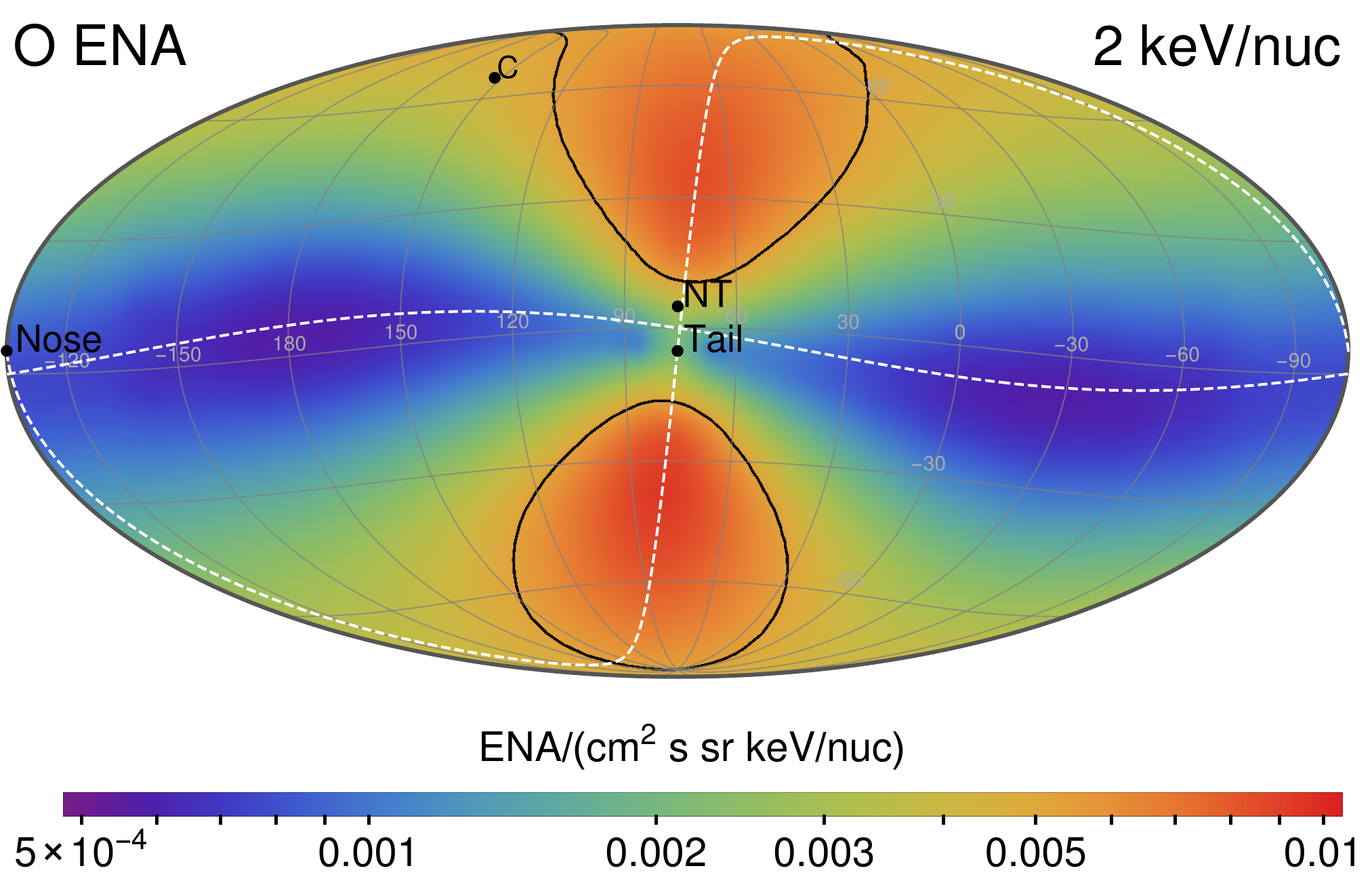}
  	\plottwo{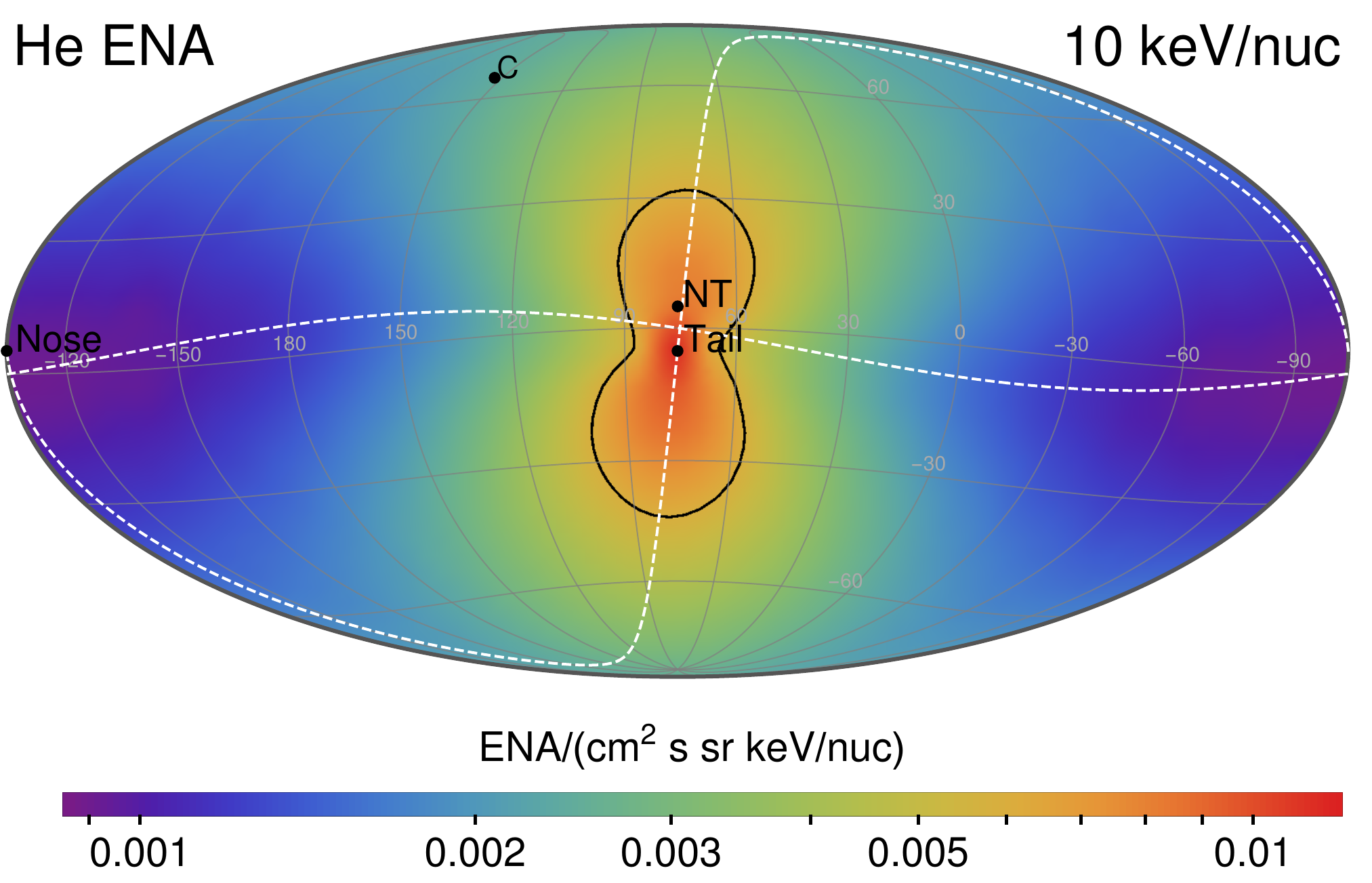}{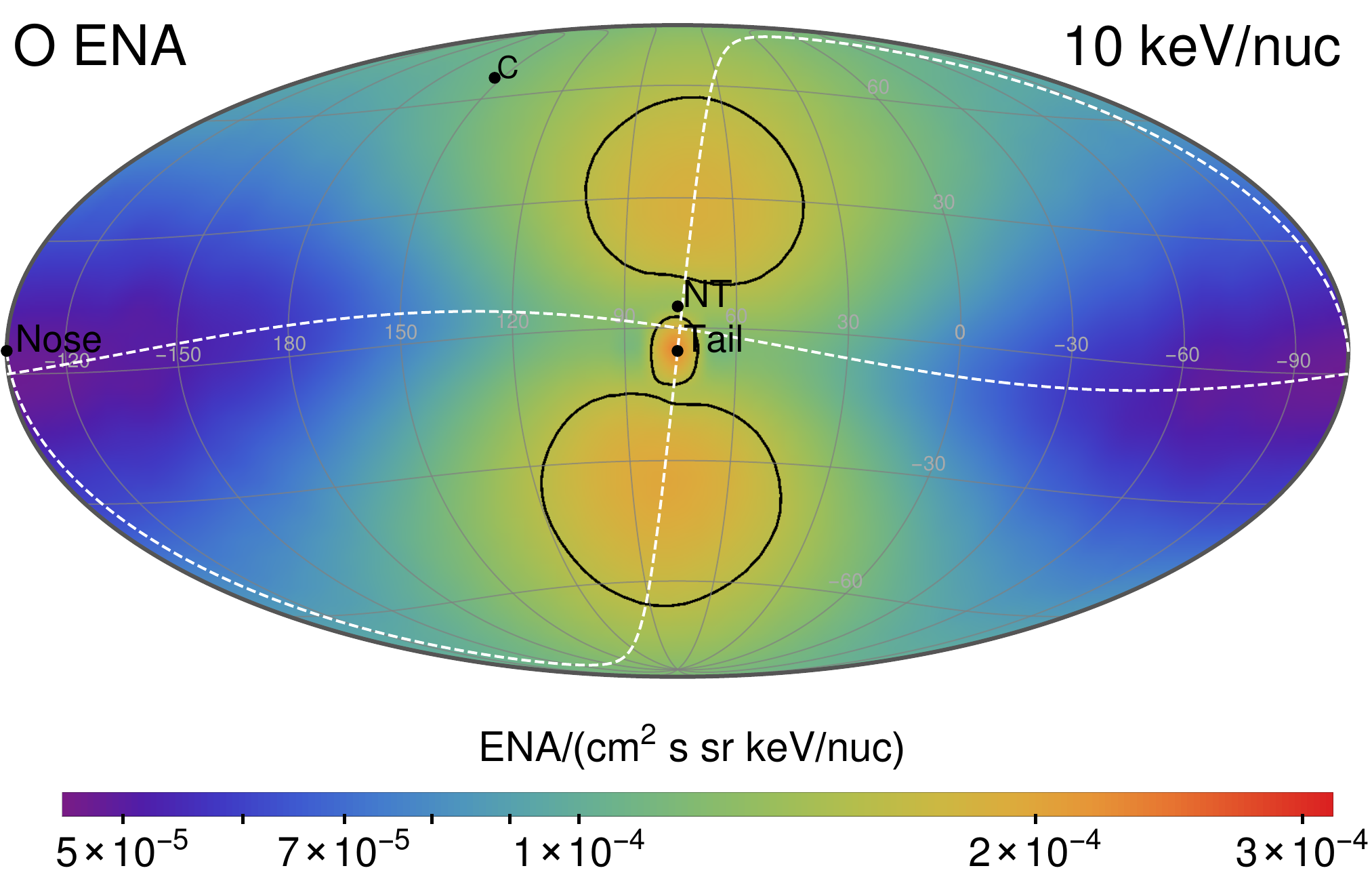}
  	\plottwo{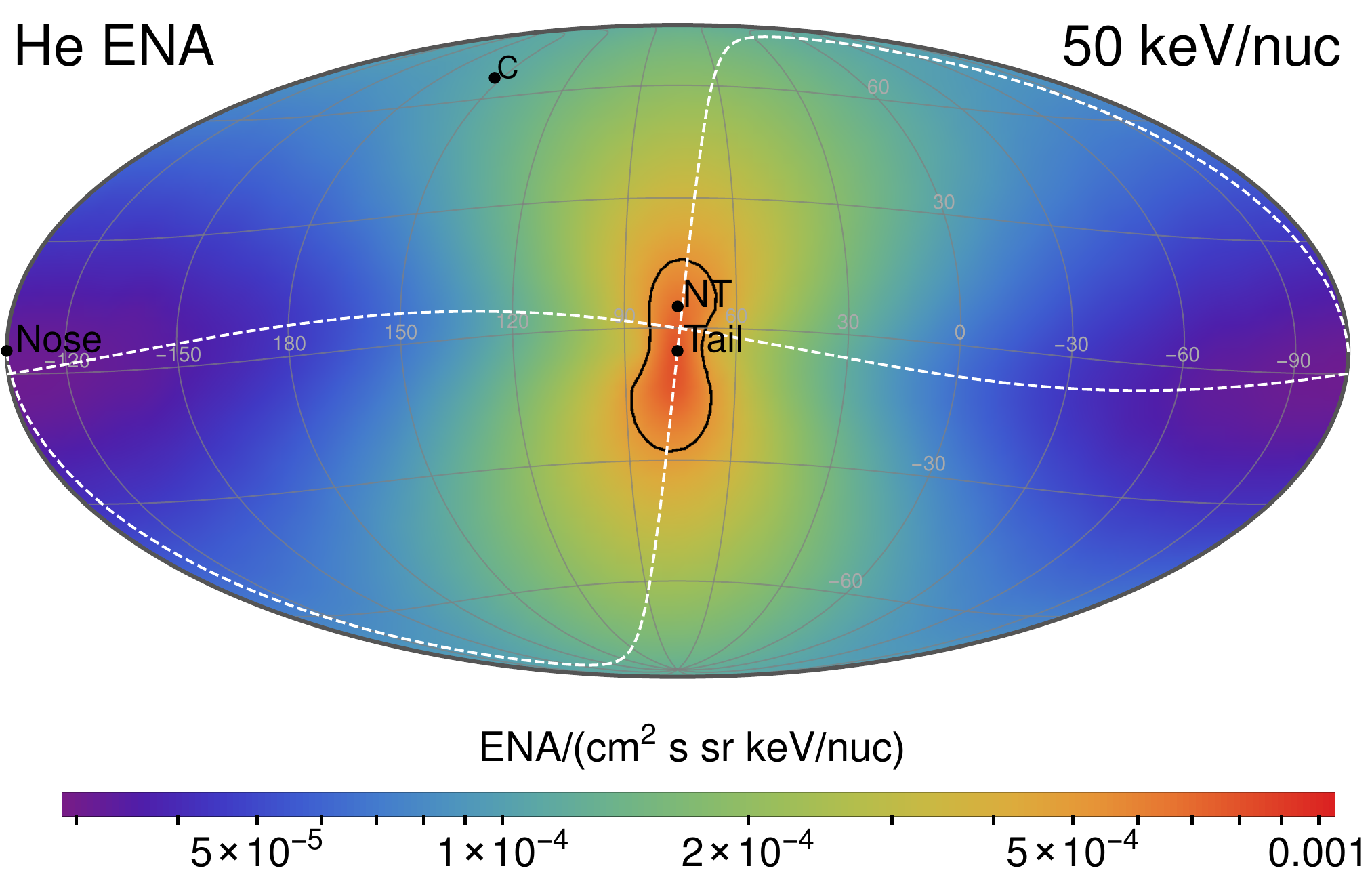}{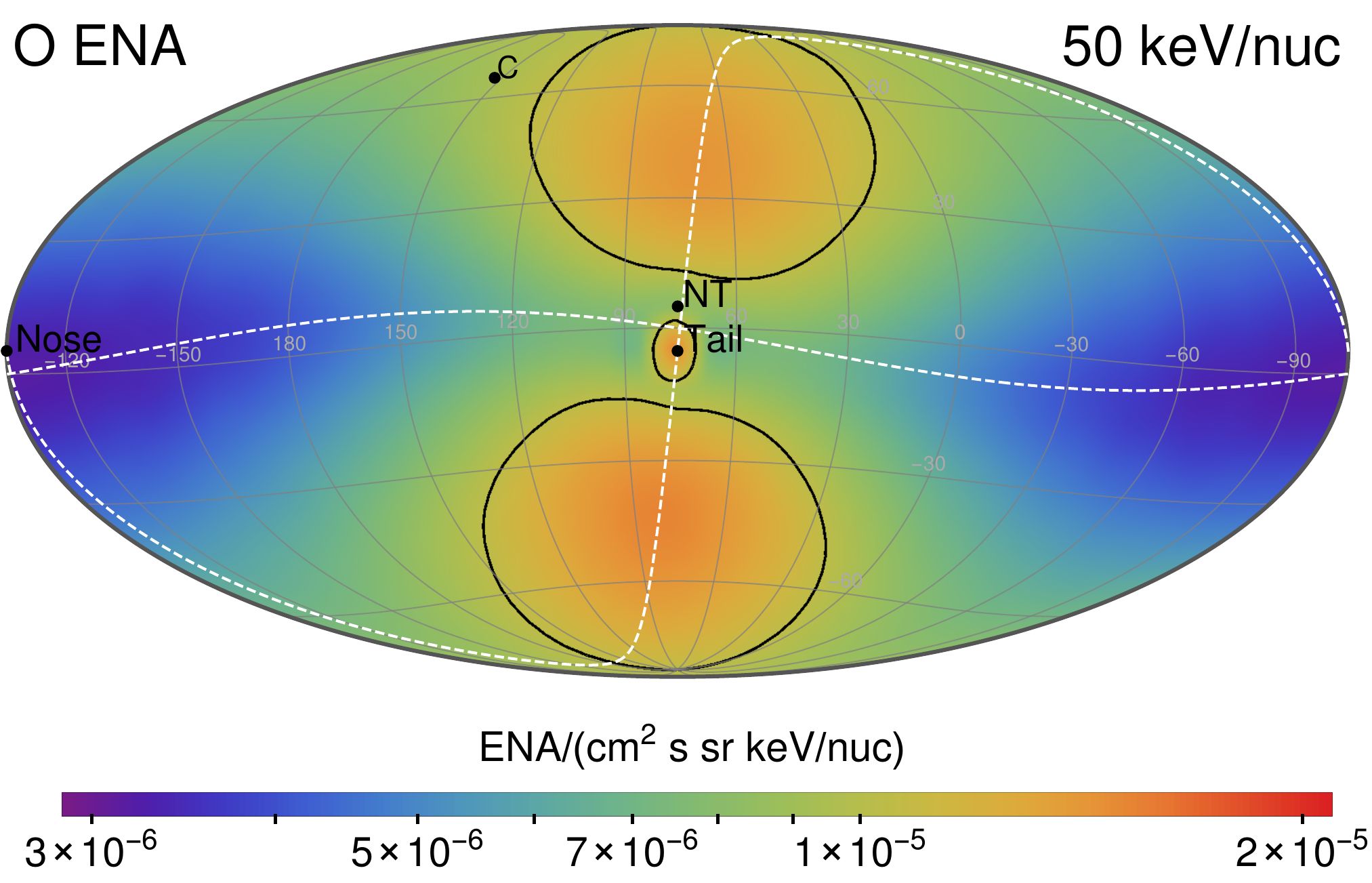}
  	\caption{Maps of the expected ENA intensities for helium (left column) and oxygen (right column) at selected energies (top-to-bottom): 0.5, 2, 10, and 50 keV/nuc. The directions selected for spectra presentation in Figure~\ref{fig:spectra} are marked with corresponding labels. Additionally, the heliographic equator and the meridian that crosses the downwind direction are marked with the white dashed lines. The black contour marks the directions for which the intensity is equal half of the maximum for each energy.\label{fig:mapsHeO}}
  \end{figure*}

\begin{figure*}
  	\epsscale{0.96}
  	\plottwo{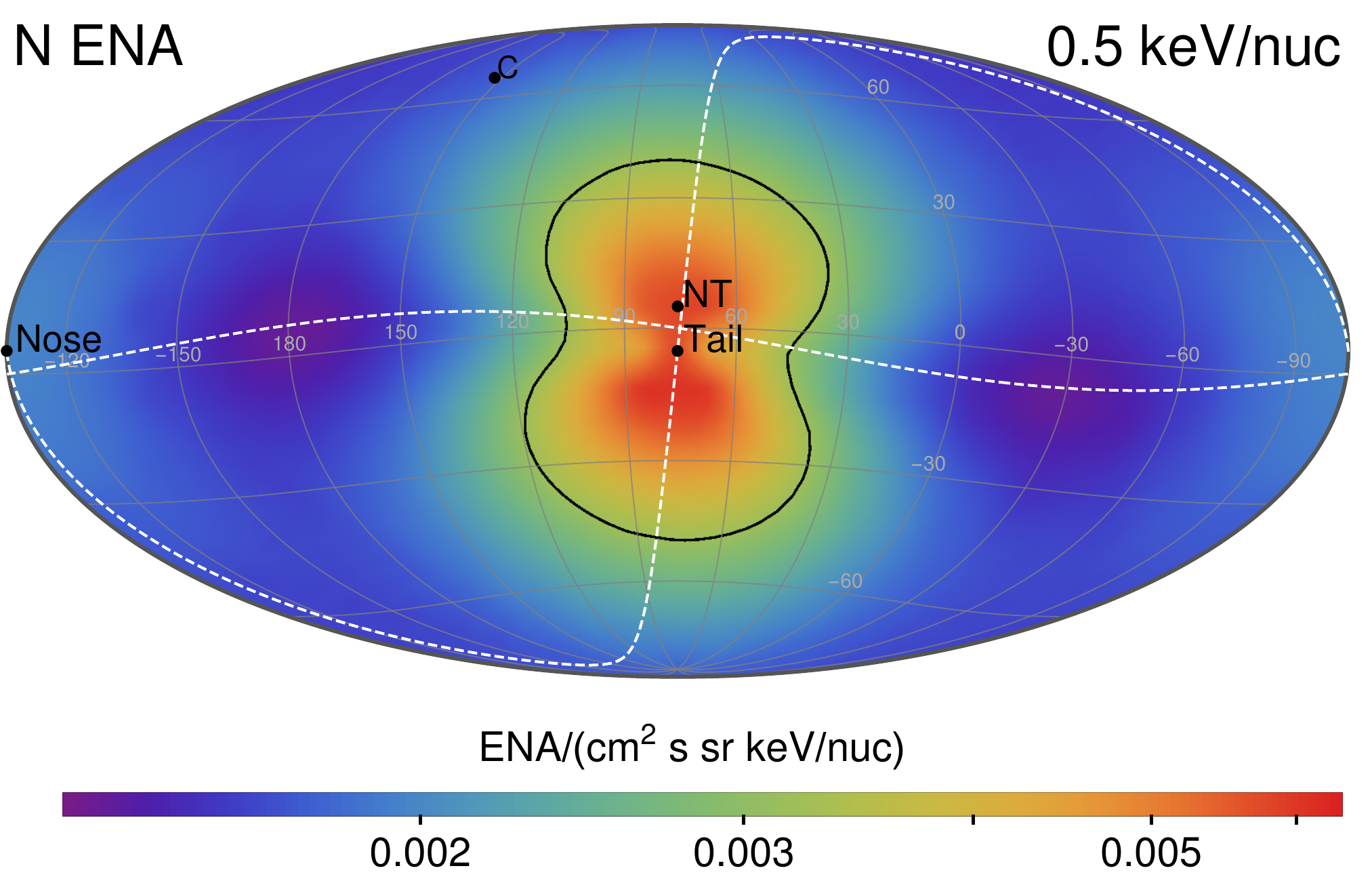}{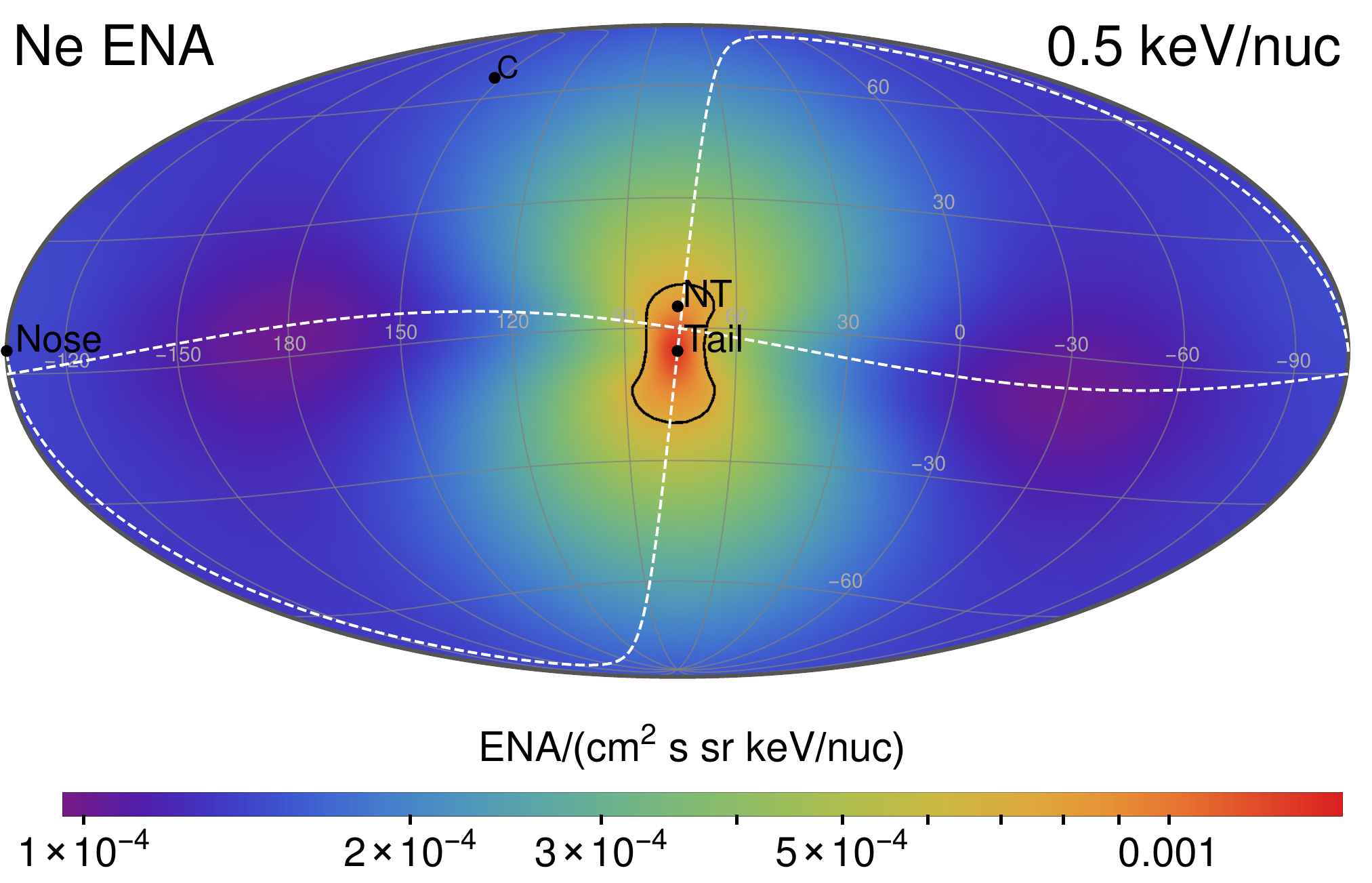}
  	\plottwo{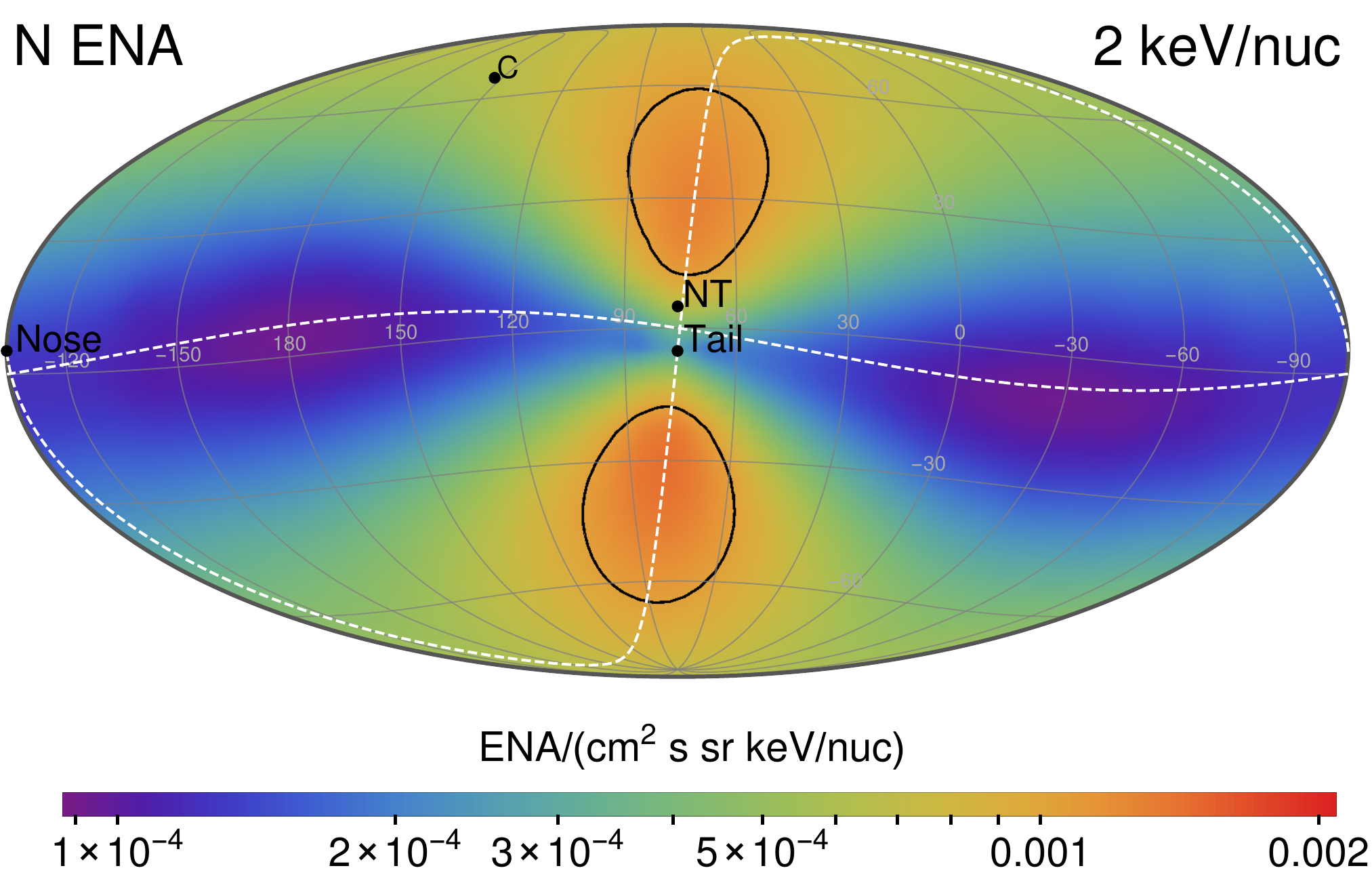}{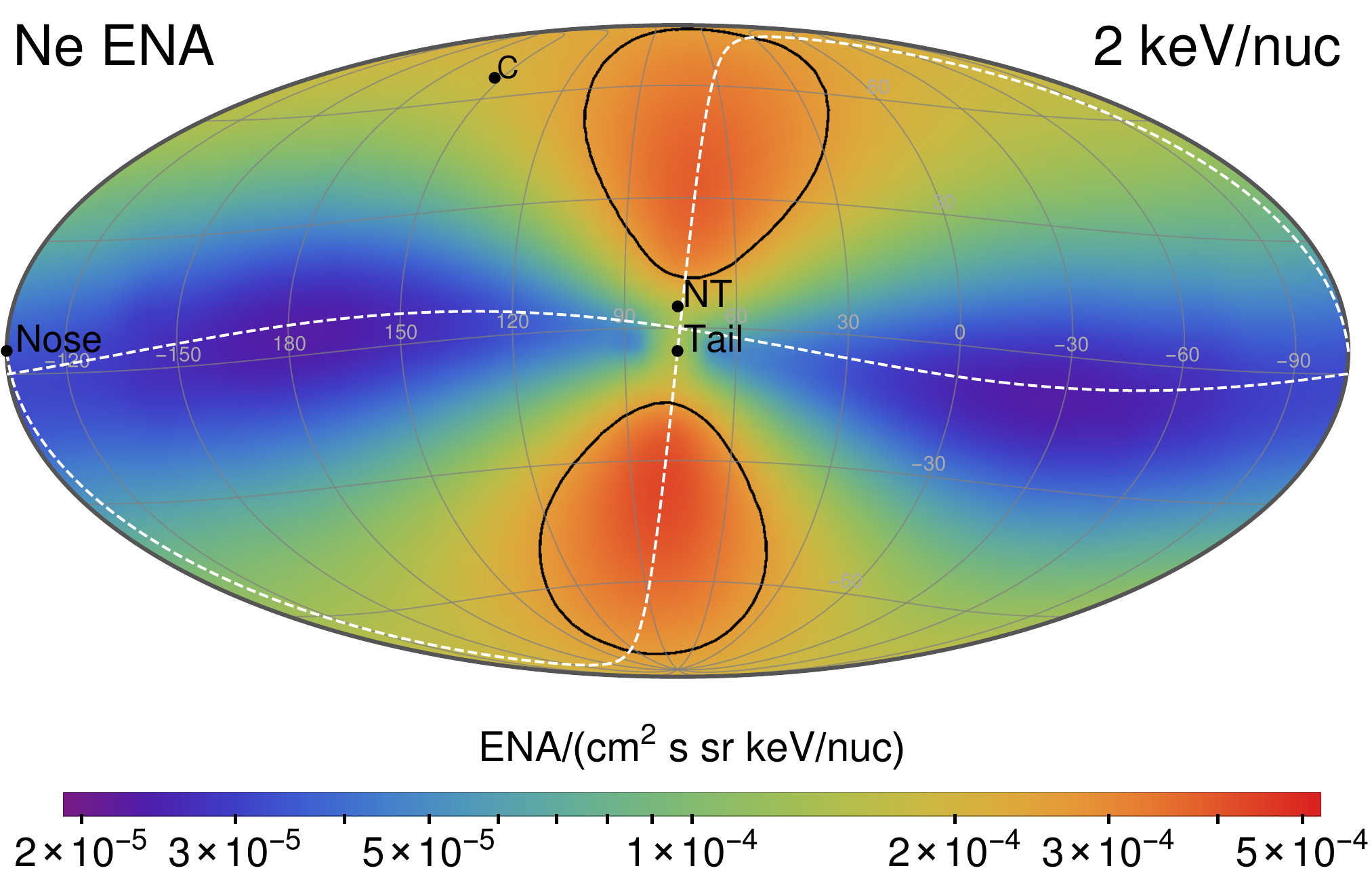}
  	\plottwo{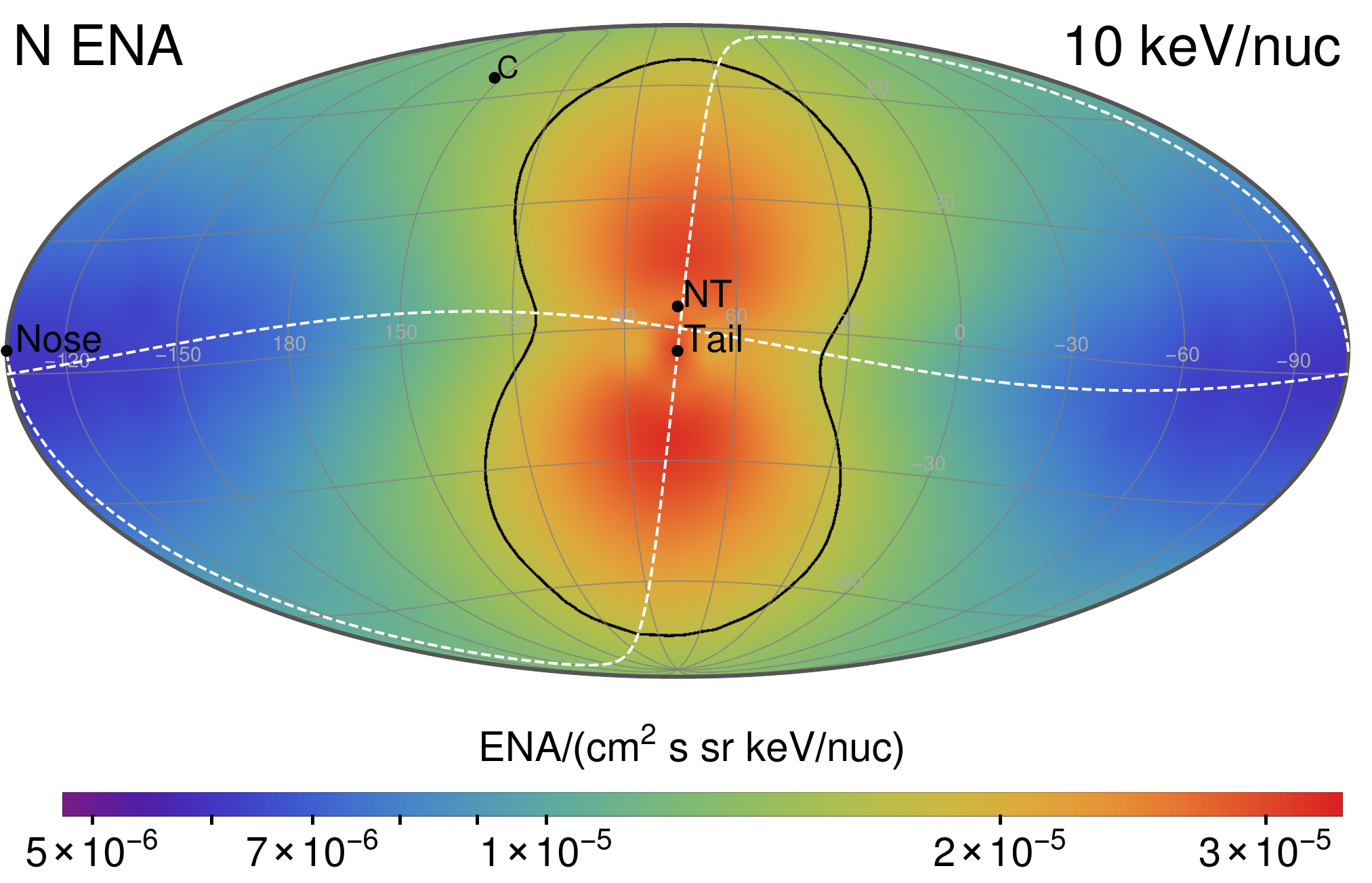}{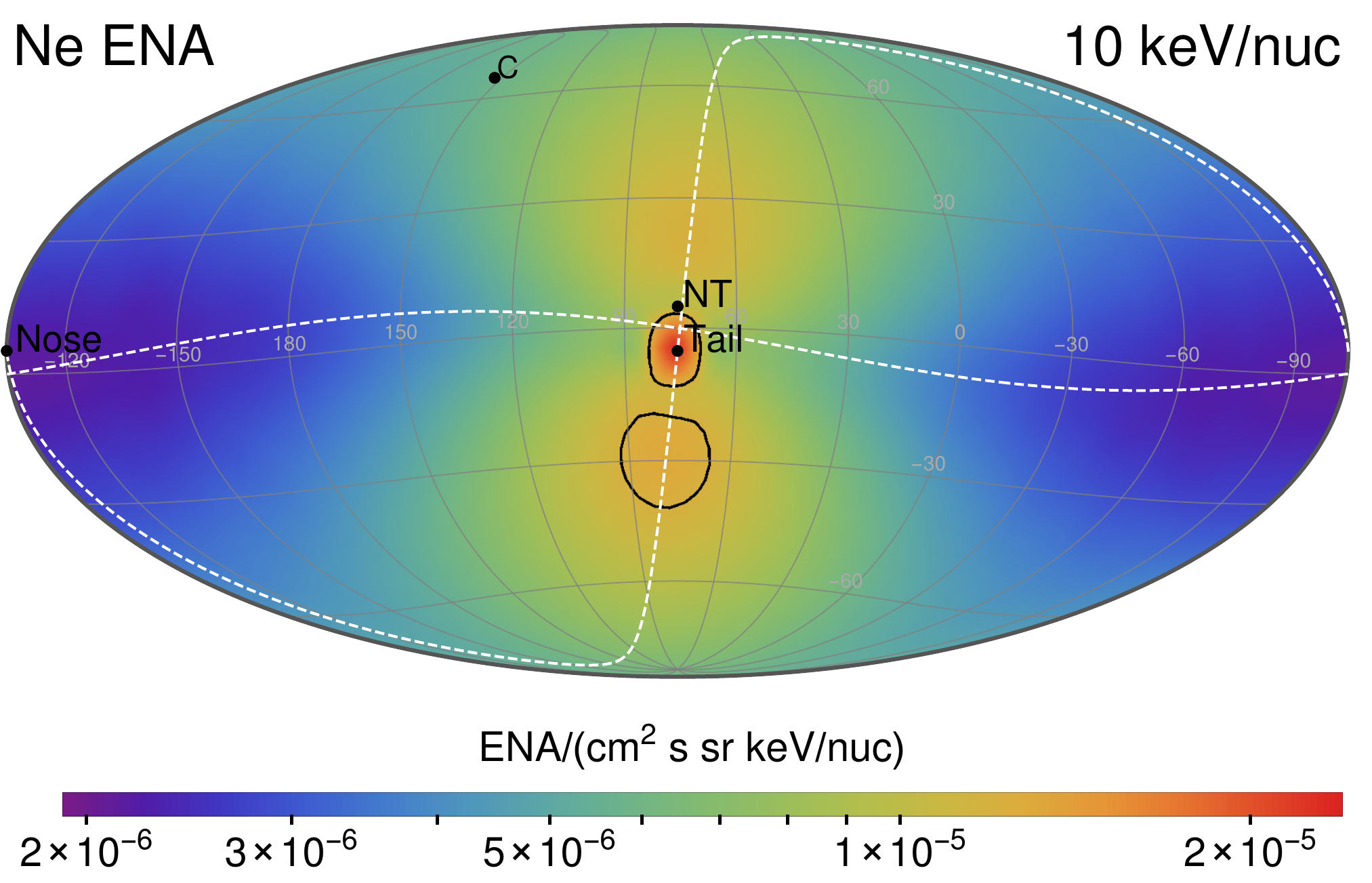}
  	\plottwo{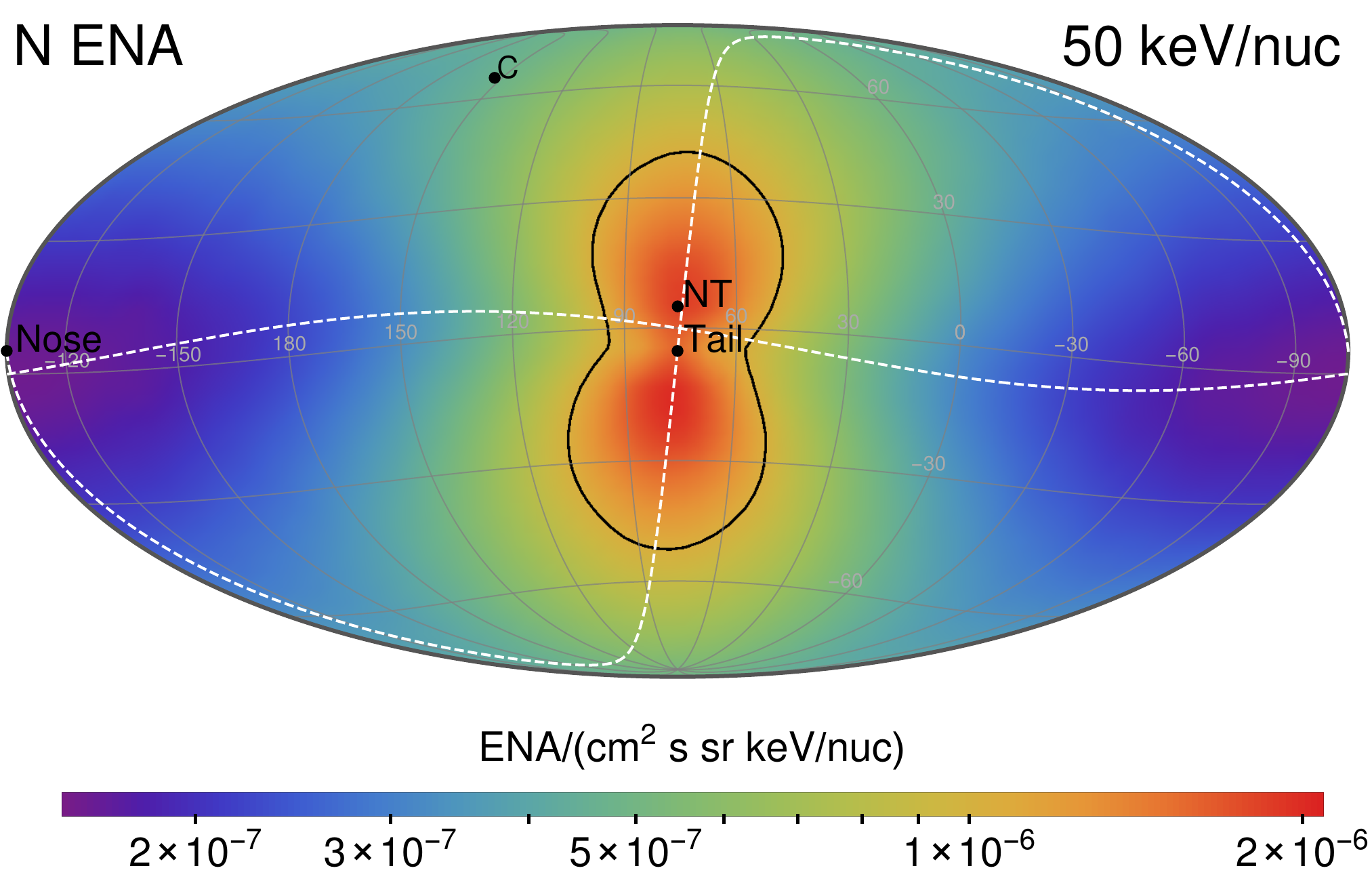}{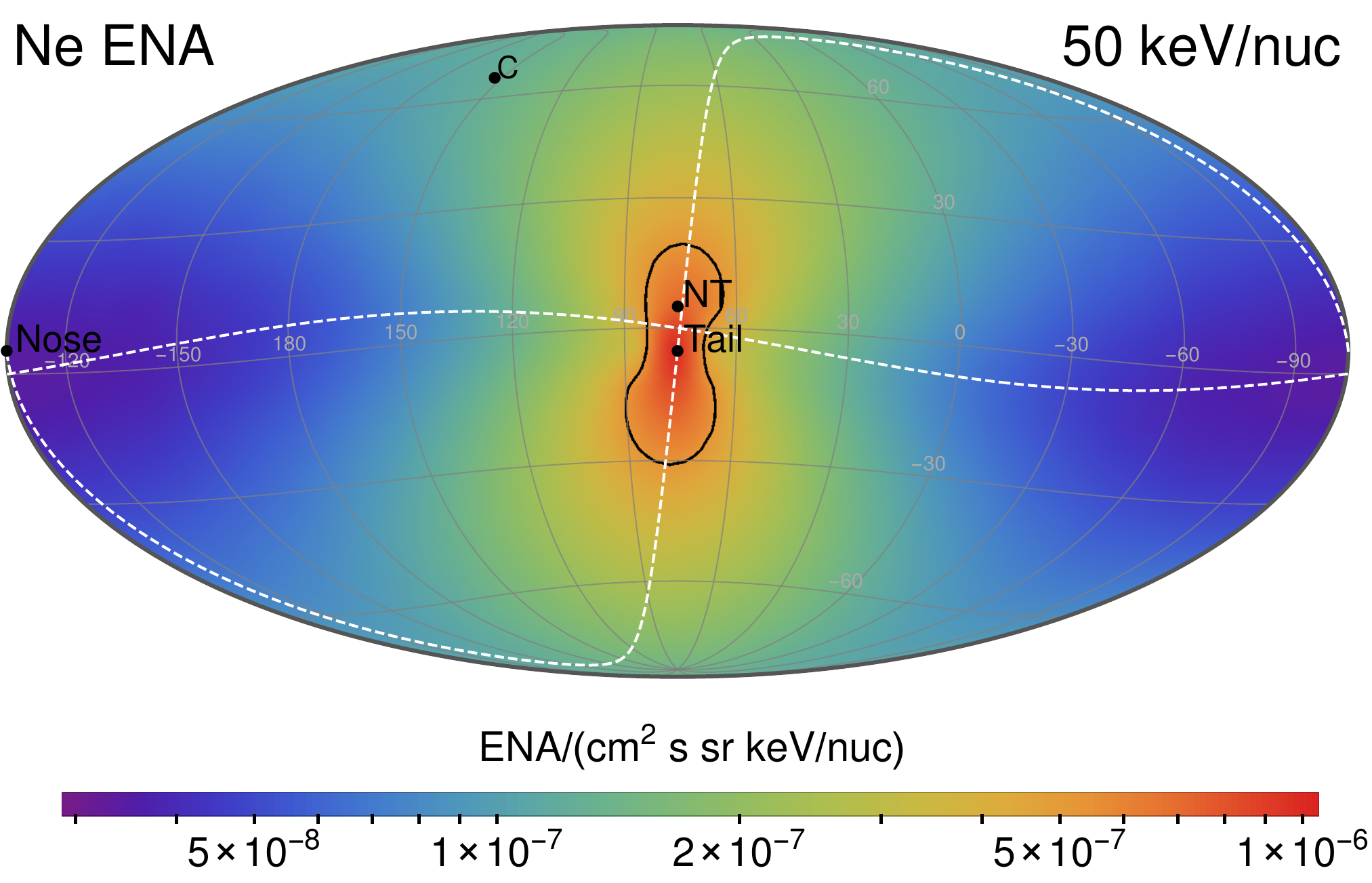}
  	\caption{Same as in Figure~\ref{fig:mapsHeO} but for nitrogen (left column) and neon (right column). \label{fig:mapsNNe}}
  \end{figure*}

Figure~\ref{fig:mapsHeO} shows the maps of expected ENA intensities for helium and oxygen at selected energies 0.5, 2, 10, and 50~keV/nuc. Figure~\ref{fig:mapsNNe} shows the maps for nitrogen and neon. The maps are centered in the downwind direction. 
 
The structure of the ENA signal in the sky is similar to that discussed for the modeled helium ENA signal in Paper I. The largest signal is expected in the downwind hemisphere and, depending on energy, the position of the maximum signal can be shifted from the downwind direction toward the Sun's poles. The signal is approximately symmetric with respect to the projection of the solar equator. However, the separation of the maxima of the signal in the south and north is generally larger for heavier atoms than for helium. 

For higher energies, the maps show also additional peaks located exactly in the downwind direction. It is a result of gravitational focusing of the heavy interstellar neutrals, which consequently create a focusing cone in the downwind direction. This subsequently enhances the PUIs fluxes at the termination shock by a factor of 2--3 in this direction. The typical FWHW of this effect is small $\sim$10$\degr$. Consequently, the resulting ENA maps can reveal this effect, but it is reduced for the smaller energies due to the Compton-Getting effect. For hydrogen ENAs, this effect is absent because due to radiation pressure interstellar neutral hydrogen does not form a focusing cone.

\subsection{Possibility of detection}\label{sec:discussion_detection}
Detectors of ENAs may be generally constructed using various designs with different methods of detection \citep{gruntman_1997}. One of the methods to recognize chemical elements is based on simultaneous measurements of full energy and velocity of the observed particle. \emph{IBEX}-Hi uses thin carbon foil to ionize the incoming ENAs and then the energies of these ionized ENAs are selected for measurements using the electrostatic analyzer \citep{funsten_2009}. However, the instrument does not allow for recognition of chemical elements. 

\begin{figure}
	\epsscale{.6}
	\plotone{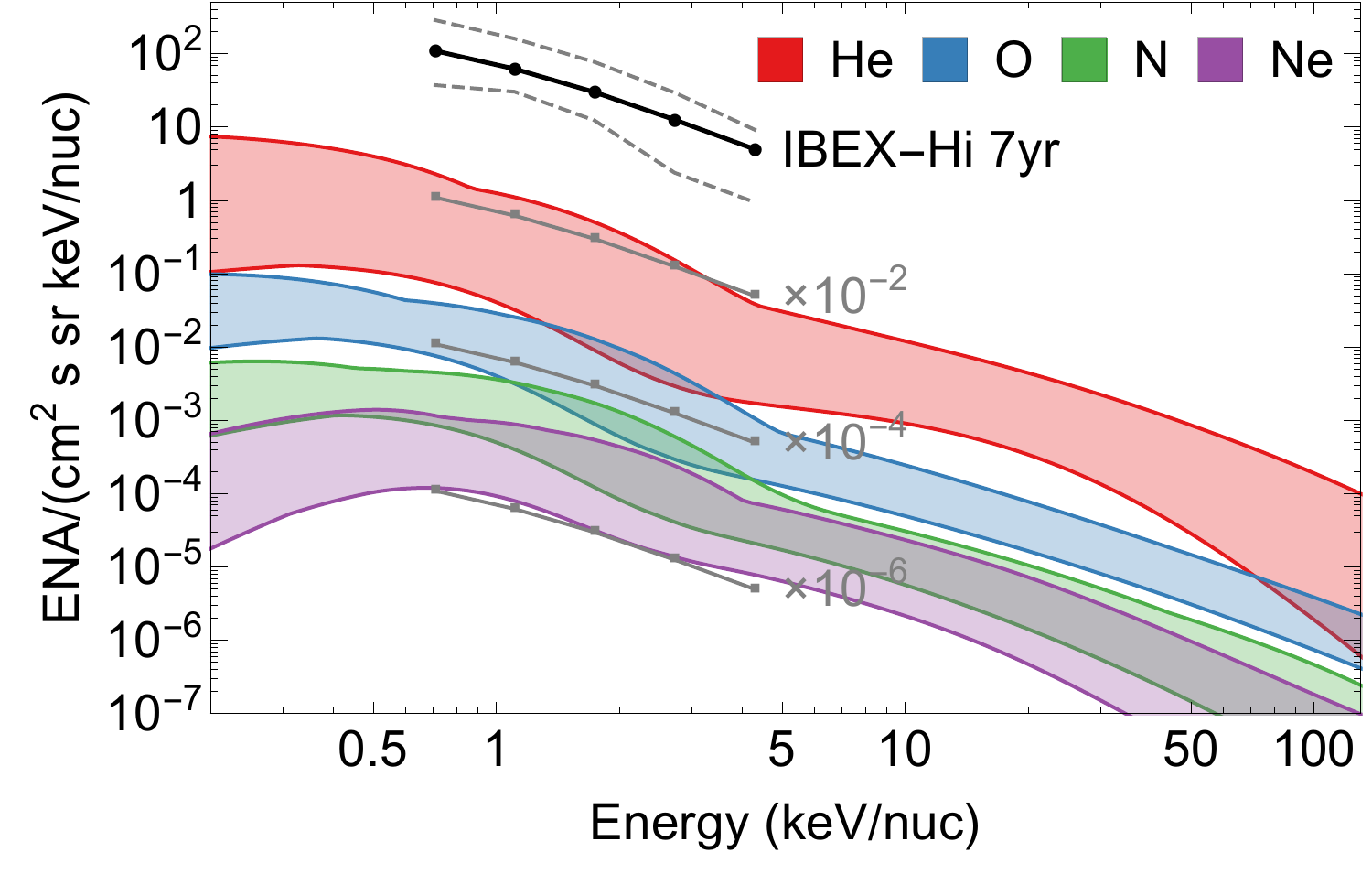}
	\plotone{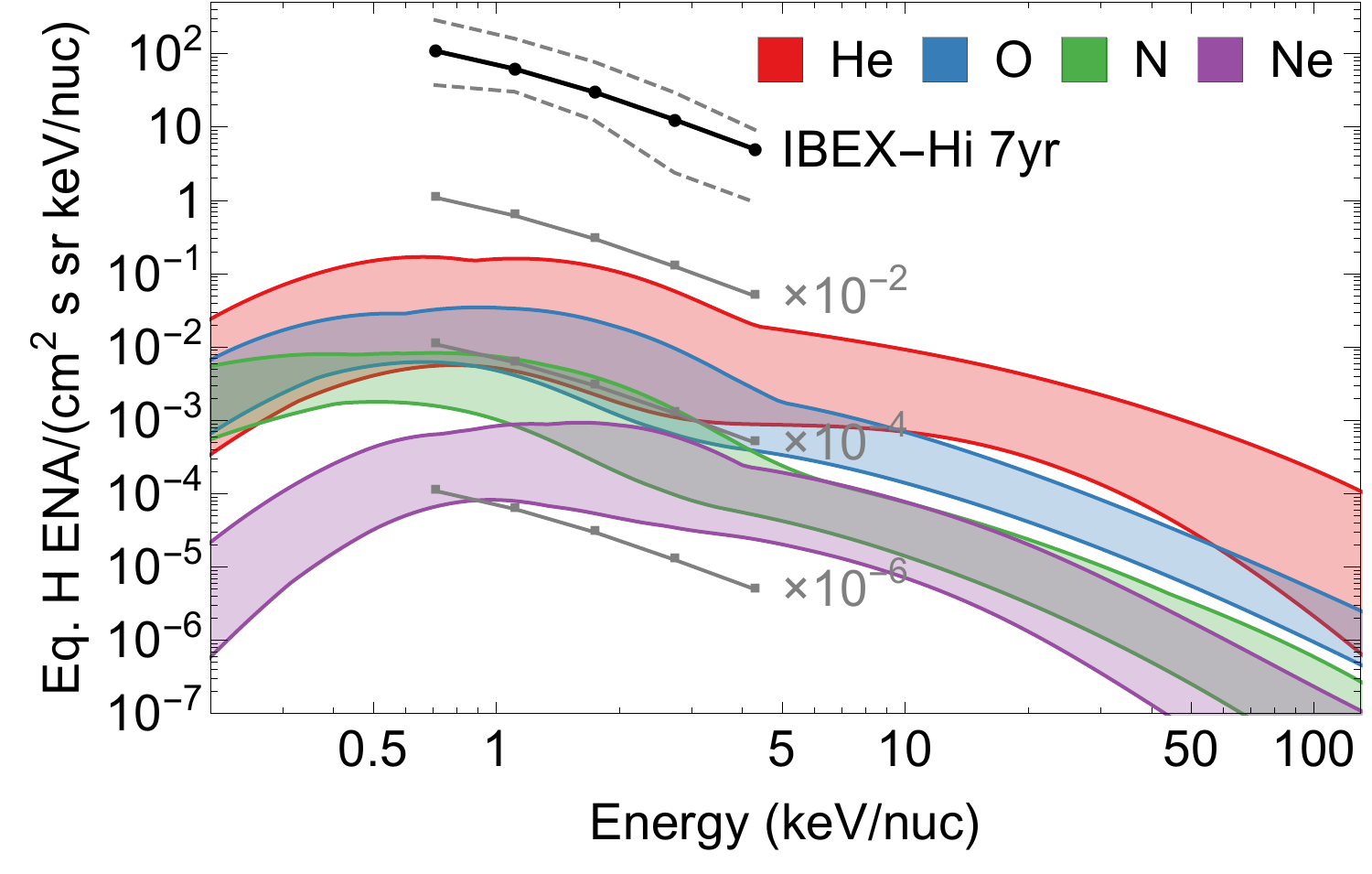}
	\caption{\emph{Upper panel:} the range of intensities of ENAs for the considered chemical elements (color bands) compared with the mean spectrum of hydrogen ENAs observed by \emph{IBEX} (black line with dots). \emph{Lower panel:} the intensities modified by the probability ratio of ionization of the considered species on the carbon foils to the ionization probability for hydrogen. The dashed gray lines show the range of the intensities observed by \emph{IBEX}. The gray lines with dots show the mean intensity observed by \emph{IBEX} multiplied by $10^{-2}$, $10^{-4}$, and $10^{-6}$. \label{fig:flux_scaled}}
\end{figure}

Carbon foils are frequently used in space instruments \citep{allegrini_2016b}. Figure~\ref{fig:flux_scaled} presents how the probability of ionization of atoms on carbon foils impacts the chance of detection of heavy ENAs. The upper panel of the figure compares the resulting ENA intensities of different species with the mean spectrum of hydrogen ENAs observed by \emph{IBEX} based on seven years of observations \citep{mccomas_2017}. The lower panel shows the same intensities multiplied with the probability ratios of ionization for the considered chemical elements to the ionization probability of hydrogen derived in Appendix~\ref{app:cfoils}. 

The expected intensities of helium ENAs in the range of energies per nucleon corresponding to that observed by \emph{IBEX} are typically 50-4000 times smaller than the observed intensities of hydrogen ENAs. For oxygen, this ratio is about $\sim$10$^{4}$, for nitrogen $\sim$10$^{5}$, and for neon $\sim$10$^{5}$--10$^{6}$. If the ionization probability on carbon foils is included, the situation slightly changes. Namely, for helium, the ratio of ionized helium ENAs to the ionized hydrogen ENAs is typically $\sim$10$^{-3}$. For other chemical elements, the ratios do not change significantly because the probabilities of ionization are comparable  for the considered species to that of hydrogen. 

The typical number of counts per $6\degr \times 6\degr$ pixel in each of the \emph{IBEX} energy channels is of the order of a few thousand after seven years of observations \citep{mccomas_2017}. The mean numbers of counts per pixel range from $\sim$1800 (for energy step 0.7 keV) to $\sim$4400 (4.1 keV), but varies substantially from pixel to pixel. Hence, observations of helium and oxygen ENAs are plausible with future ENA detectors, for which a significant increase of efficiency is expected. Observation of nitrogen and neon is less likely due to smaller intensities expected from the inner heliosheath.

This reasoning neglects the efficiency of particle detection. In \emph{IBEX}-Hi this detection is possible due to secondary electron emission caused by passing of a particle through two additional carbon foils \citep{funsten_2009a}. However, the yields of electrons emitted is less species-dependent, and it is mostly similar for the corresponding energy per nucleon \citep{allegrini_2016b}. 

Nonetheless, the expected number of counts based on this simple estimation suggests that the number of events related to heavy ENAs compared to hydrogen ENAs could be less by a factor of 200 to a few thousand. This means that the crucial point for detection of heavy ENAs is to have a very small background from non-ENA events. Additionally, the mass spectrometer must have a sufficient resolution, allowing for recognition of heavy atoms without false classification of hydrogen ENAs.  

\section{Discussion}\label{sec:discussion}
The structure of the intensities of heavy ENAs presented in this analysis is a result of two effects. The first one is the variation of the mean PUI energies with heliographic latitude. For larger latitudes, this energy is more than two times larger than close to the equator (Section~\ref{sec:PUIdistribution}). The other effect is due to the mean free path of heavy ions in the inner heliosheath (Appendix~\ref{app:cxih}). Consequently, the shape of the observed signal in the downwind direction resembles an hourglass for some energies. This finding was formulated in Paper I for helium, but remains valid also for the other considered chemical elements. 

\begin{figure}
	\epsscale{1}
	\plottwo{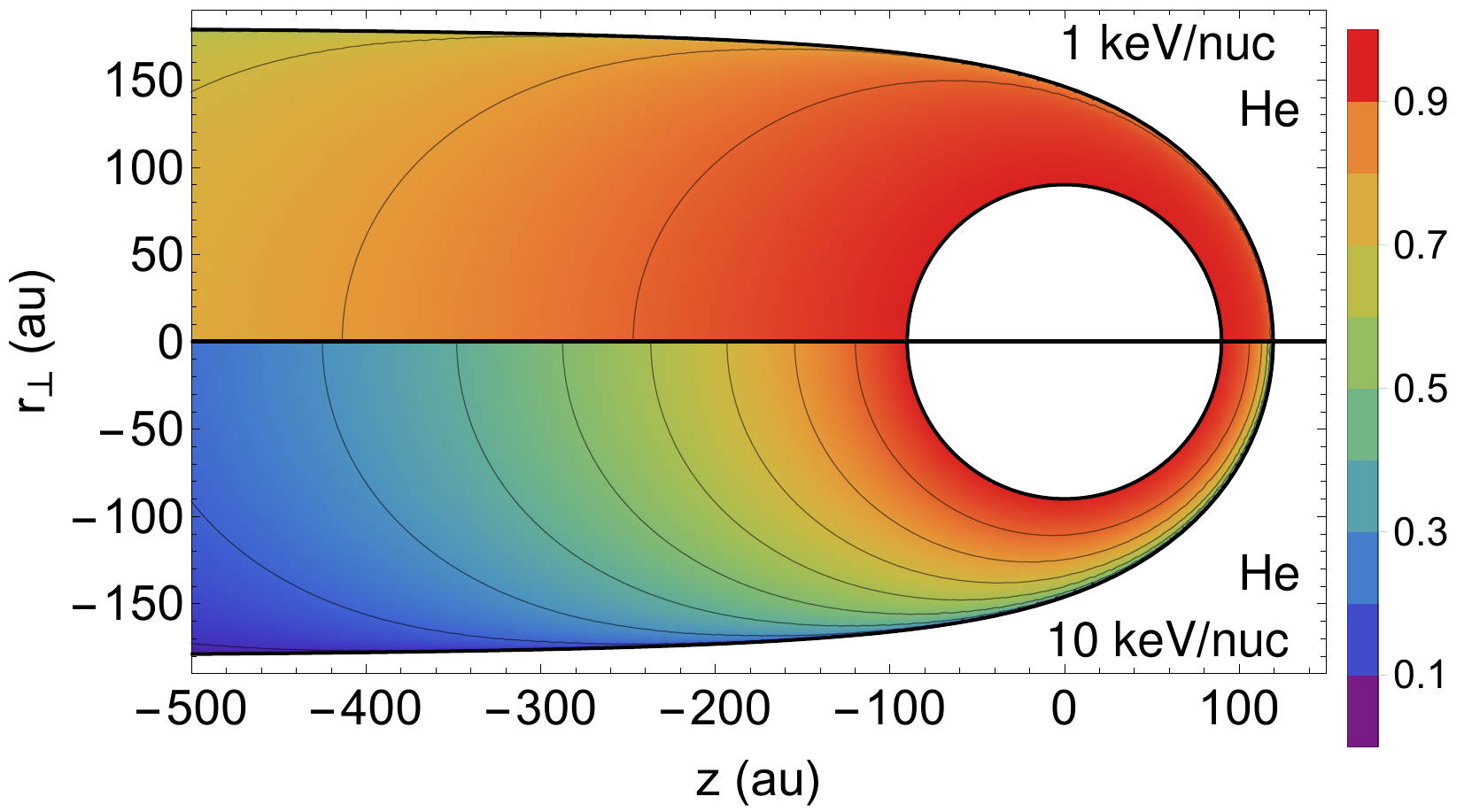}{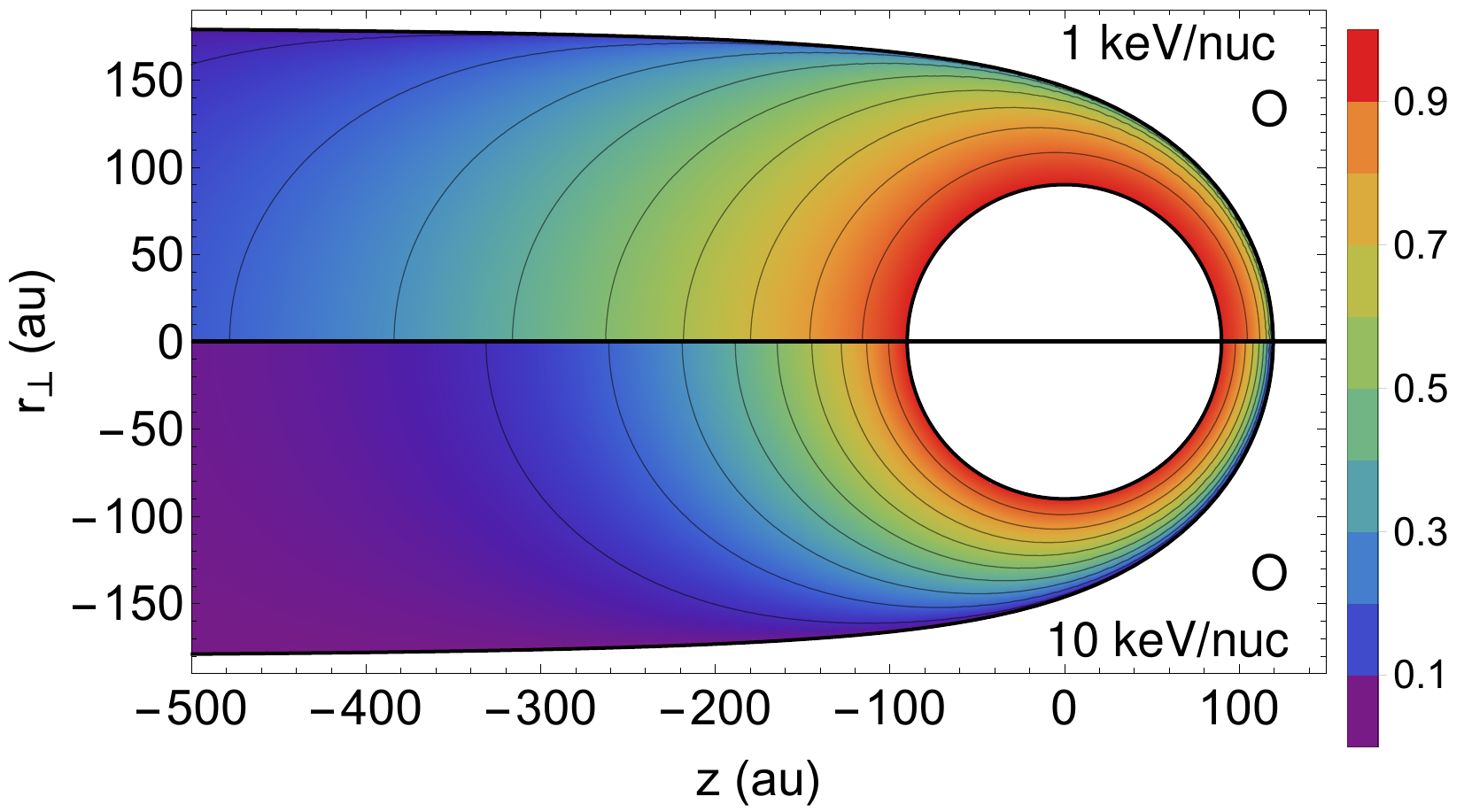}
	\plottwo{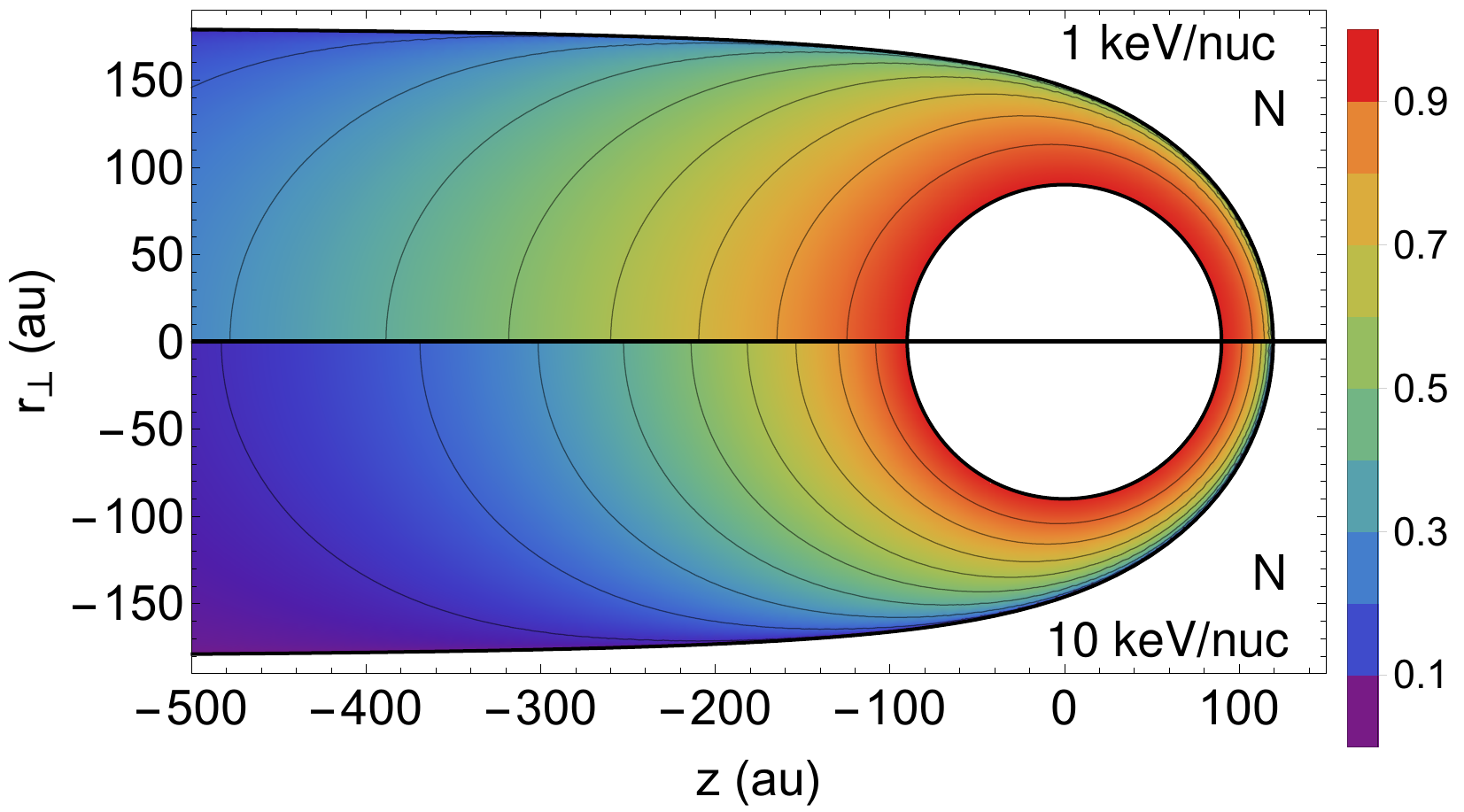}{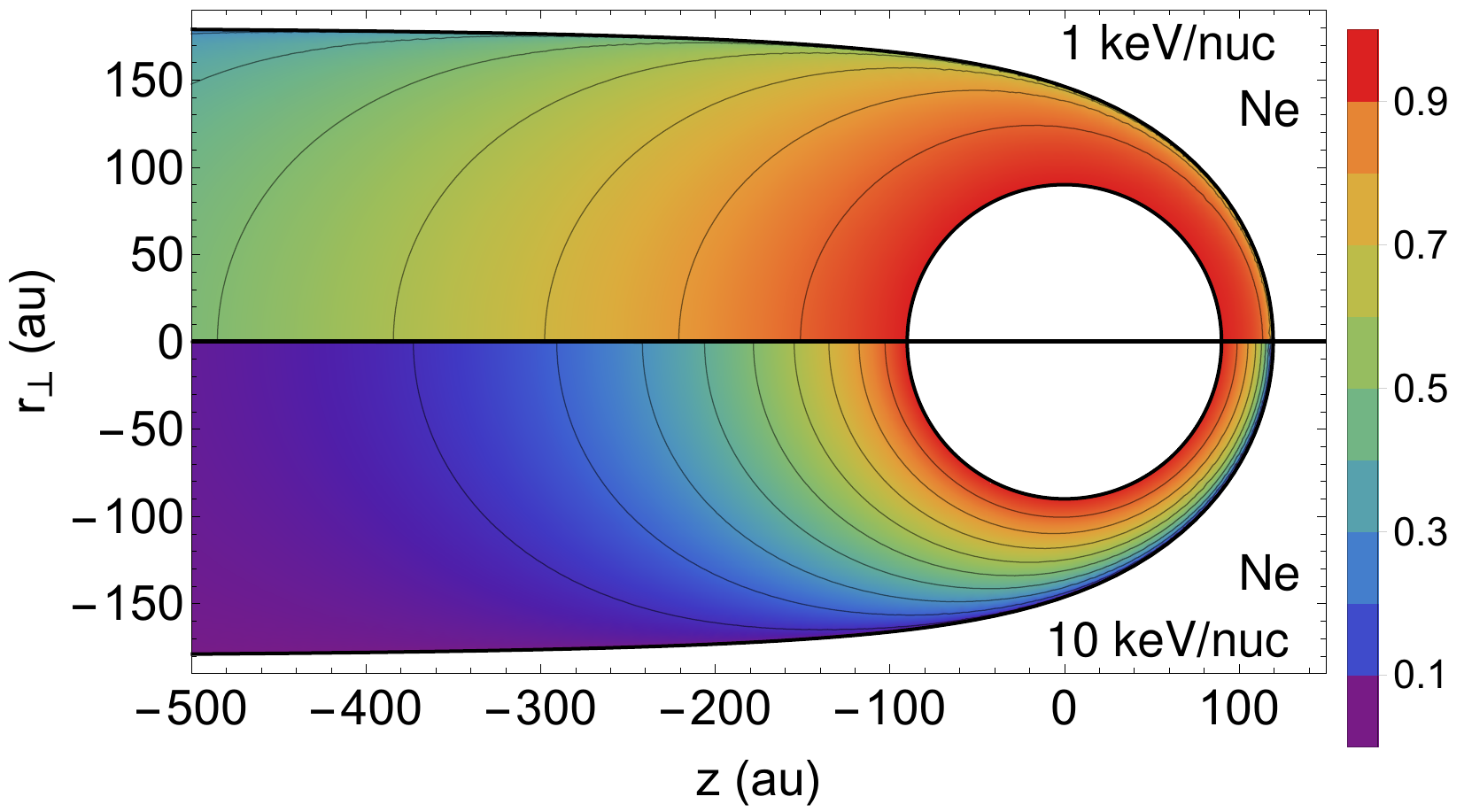}
	\caption{Probability that heavy ions are not neutralized due to charge exchange with the interstellar neutrals at their paths from the termination shock. The panels show results for helium (upper-left), oxygen (upper-right), nitrogen (lower-left), and neon (lower-right). The upper and lower parts of each panel show the probabilities for ions of energy 1 keV/nuc and 10 keV/nuc, respectively. \label{fig:surv}}
\end{figure}

The differences in the cross-sections for the charge exchange of heavy PUIs are reflected in the ion penetration depths in the inner heliosheath. Figure~\ref{fig:surv} presents probabilities that heavy ions are not neutralized at their paths from the termination shock. With an increasing rate of charge-exchange reaction in the inner heliosheath, which is a product of the corresponding cross-section  (Appendix~\ref{app:cxih}) and velocity, the penetration depth of ions decreases. Due to lower concentration in the tail of the ions that are more effectively neutralized, the tail-to-nose ratio of ENA intensities is then smaller, because there are fewer ions in the tail available for ENA production.

Due to larger cross-sections for the charge exchange of PUIs for oxygen, nitrogen, and neon (Appendix~\ref{app:cxih}) than it is for helium, the signal is less concentrated in the downwind direction. This effect can be easily noticed from the black contours shown in Figures~\ref{fig:mapsHeO} and~\ref{fig:mapsNNe}. For helium, the tail-to-nose ratio of ENA intensities is very large, thus the black contours, which mark half of the maximal signal, are confined very close to the tail. For oxygen, much larger parts of the sky have the signal larger than half of the maximum. Moreover, the positions of the maxima located away from the downwind direction are much more separated. In Paper I, we found that the typical separation between the southern and northern maxima is 10$\degr$--15$\degr$. For oxygen, this separation can reach $\sim$80$\degr$. The situation is similar for nitrogen. 

For neon, the used cross sections are similar to those of oxygen and nitrogen for energies of $\sim$2--10~keV/nuc, but they are noticeably smaller outside of this range. For 2~keV/nuc, the cross-section for neutralization of nitrogen and neon PUIs is almost the same, and in fact the obtained maps are similar in the terms of relative intensities. In this analysis, the shape of the spectra of PUIs at the termination shock are assumed proportional (they differ only due to the densities of PUIs). Hence, potentially observed differences in maps of the relative ENA intensities will likely indicate differences in the energization processes at the termination shock. However, differences due to the details of the measurement process, discussed in the preceding section, must be appropriately taken into account.

The maps of oxygen and nitrogen show that the region embedded by the black contour split for energy 2 keV/nuc, but merge for energies 0.5 and~10 keV/nuc. This behavior is similar to that noticed in the observations of hydrogen ENAs \citep{zirnstein_2016, zirnstein_2017}. 

This analysis is based on a simple model of the potential flow of incompressible plasma in the inner heliosheath. The result may slightly change for more advanced models. For instance, in the models with the interstellar magnetic field included, the heliosphere is significantly distorted. In this situation, the integration path can change. In particular, if the tail is deflected from the downwind direction, also the hourglass-shaped feature should be significantly deflected.

\section{Summary}\label{sec:summary}
Prospects for observations of heavy ENAs are presented in this paper. The expected intensities were calculated based on modification of the model proposed by \citet{grzedzielski_2013, grzedzielski_2014} and then extended in Paper I. The model is described in Section~\ref{sec:method}. In the model, ENAs result from neutralization of PUIs in the inner heliosheath. The densities of PUIs were calculated from assessing the source intensity due to ionization of the interstellar neutrals in the supersonic solar wind through photoionization and charge-exchange processes. We considered the four most abundant heavy species in the neutral component of the interstellar medium: helium, oxygen, nitrogen, and neon. The spectra of PUIs were assumed to follow the same distribution functions in the velocity space but with different densities at the termination shock. This assumption was found to be supported by \emph{Voyager 1} observations. These distribution functions were evolved in the inner heliosheath due to charge-exchange processes with the interstellar neutrals. The same processes are responsible for the creation of the ENAs.

Expected intensities of heavy ENAs as spectra in some selected directions and maps for energies 0.5, 2, 10, 50 keV/nuc are presented in Section~\ref{sec:results}. The widest range of these intensities is expected for helium, for which the tail-to-nose ratio is $\sim$50 for energies $\sim$3~keV/nuc. All considered species show large tail-to-nose intensity ratios, that are typically smaller for oxygen and nitrogen: 3--10, and larger for neon: 5--20. 

Possibility of detection is discussed in Section~\ref{sec:discussion_detection}. The expected intensities of helium ENAs are smaller 50--4000 times than the intensities of hydrogen ENAs observed by \emph{IBEX}-Hi for the corresponding range of energy per nucleon. For oxygen, this factor ranges from $2\times 10^3$ to $4\times 10^4$, for nitrogen: $1.5\times 10^4$ to $3\times 10^5$, and for neon: $5\times 10^4$ to $9\times 10^5$. Since future detectors of ENAs are likely to be equipped with carbon foils necessary for the ionization of ENAs, the probabilities of ionization on these foils should also be taken into account. For helium, this probability is 10 times smaller than that for hydrogen for energy 1~keV/nuc but it increases for higher energies and is almost the same for atoms with energies $\gtrsim$30 keV/nuc. For the other considered species, this probability is slightly larger (up to two to three times) than for hydrogen at energies 2--100~keV/nuc. Consequently, the observations of helium and oxygen are likely possible due to the increased efficiency expected on future instruments. Because of the smaller number of events expected for nitrogen and neon, the observation of single events may be possible, but likely not sufficient to obtain statistically significant maps of their emission. 

The long mean free paths of helium PUIs should allow for better understanding of the global structure of the heliosphere (see Paper I for more discussion). The other considered species have generally shorter mean free paths, and thus the corresponding intensities do not show large tail-to-nose ratios.

Heavy ENAs are a good target for future ENA detectors. They could bring information on global structure of the heliosphere, as well as on energization processes in the heliosphere. Successful observation requires that the detector be capable of determining the mass of the incoming atoms with a good resolution. The dominant hydrogen atoms must not be mistaken for helium with a probability much smaller than 1:10$^4$, and for a heavier atom that is even smaller than 1:10$^6$. Achieving such an accuracy would allow for the establishment of a new type of heliopshere observation through heavy ENAs. 

\acknowledgments

The authors are grateful to Stan Grzedzielski for valuable discussions and helpful comments regarding the manuscript. The research was supported by the grant 2015/19/B/ST9/01328 from the National Science Centre, Poland. P.S. is supported by the Foundation for Polish Science (FNP).

\appendix
\section{Charge exchange between neutrals and protons}\label{app:cxnp}
The charge exchange between interstellar neutral atoms and solar wind protons is one of the reactions producing PUIs in the supersonic solar wind. The bulk speeds of the solar wind in the inner heliosphere range from $\sim$300~$\mathrm{km\,s^{-1}}$ to $\sim$1000~$\mathrm{km\,s^{-1}}$. These speeds are approximately equal to the relative speed between protons and interstellar neutrals because the bulk speeds of interstellar neutrals, their thermal speeds, and the thermal speeds of solar wind protons are significantly smaller than the speed of expansion of the solar wind, at least in the supersonic solar wind. 

The cross-sections of these reactions were sought based on the earlier analyses of the modulation of interstellar neutrals and from references collected in AMBDAS\footnote{\url{https://www-amdis.iaea.org/AMBDAS/}}. For helium, oxygen, and neon the same sources are used as in recent analyses on interstellar neutrals \citep{bzowski_2013c, bzowski_2013b, sokol_2016}. The cross-section for charge exchange of neutral helium and protons is taken from \citet{barnett_1990}. The formulae from \citet{lindsay_2005} and \citet{nakai_1987} are adopted for charge exchange of oxygen and neon, respectively. According to AMBDAS, experimental measurements of charge exchange between nitrogen and proton are available only for molecular nitrogen, but not for atomic, which is needed in this analysis. Consequently, the theoretical cross-section calculated by \citet{cabrera-trujillo_2000a} is used in this analysis.

\begin{figure}
	\epsscale{.6}
	\plotone{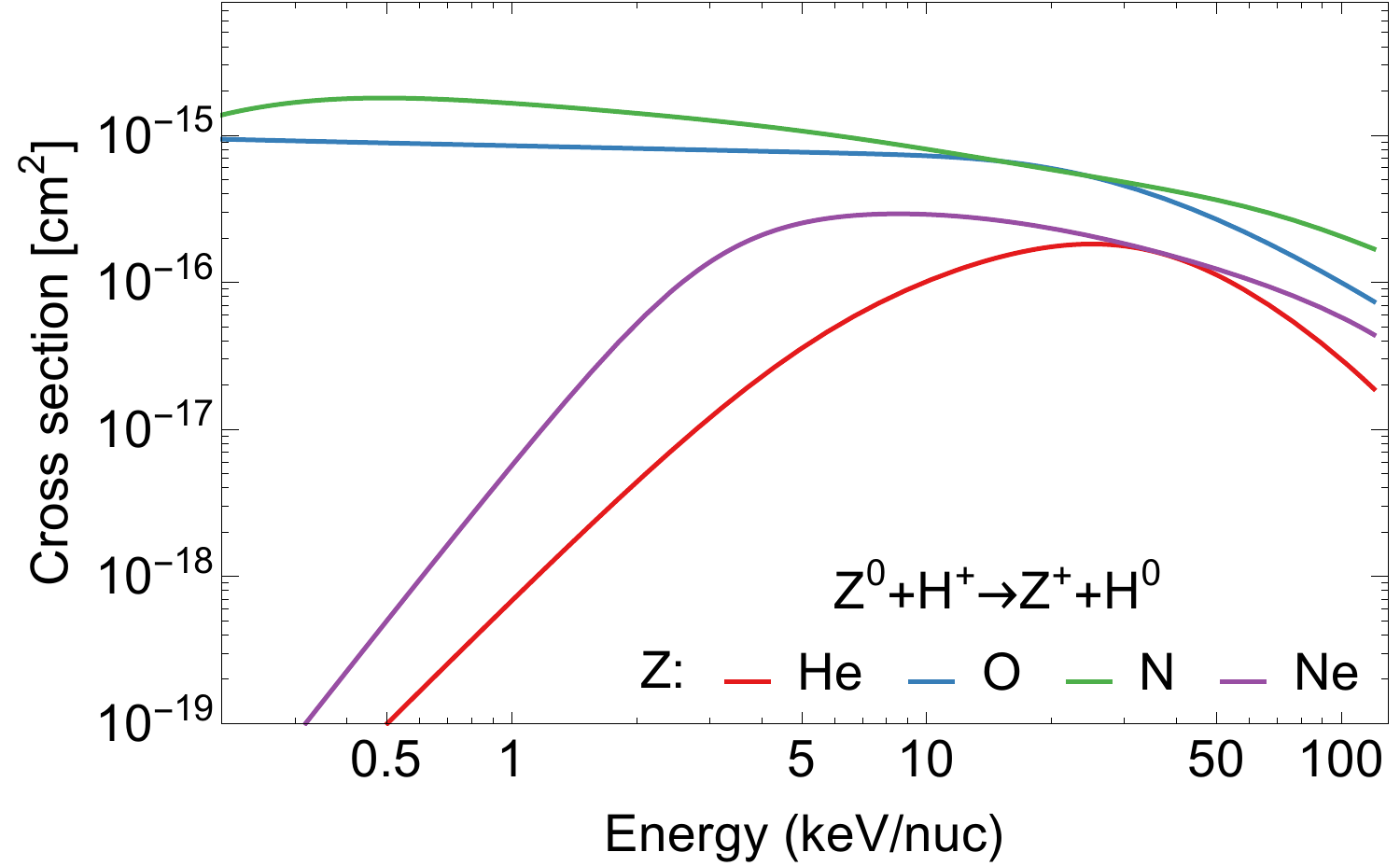}
	\caption{Cross-sections for charge exchange between neutral atoms and protons as a function of projectile energy. Colors represent the considered chemical elements. \label{fig:cxnp}}
\end{figure}

The cross-sections for the charge exchange processes considered in this appendix are presented in Figure~\ref{fig:cxnp}. For energies $\sim$1 keV/nuc, corresponding to the solar wind speeds, the cross-sections for oxygen and nitrogen are large, $\sim$10$^{-15}$~$\mathrm{cm^{2}}$, and they change only slightly with energy. The situation is different for helium and neon, for which these cross-sections are more than 100 times smaller and strongly depend on energy in the considered range. 

\section{Charge exchange of ions and neutral hydrogen}\label{app:cxih}
Neutralization of PUIs on the interstellar neutral hydrogen atoms is the dominant reaction for production of ENAs. Cross-sections for these reactions are crucial for a wide range of energies considered in this analysis.

The cross-section for charge exchange of He$^+$ ion with a hydrogen atom is adopted from \citet{barnett_1990}. For oxygen, the cross-section given by \citet{lindsay_2005} is used. There is no formula available for the charge-exchange cross-section of N$^+$ ion with neutral hydrogen atoms either from theoretical calculations or from fitting of multiple experiments. Consequently, the experimental results from \citet{phaneuf_1978a}, \citet{nutt_1979}, and \citet{stebbings_1960} are fitted to a third degree polynomial in the logarithmic domain. The used data spread over an energy range from 0.02 keV/nuc to 40 keV/nuc. The obtained fit agrees with the theoretical calculations made for energies $<$1 keV/nuc by \citet{lin_2005}. Based on AMBDAS, the cross-section for charge exchange of Ne$^+$ ion with the hydrogen atom has not been measured, and there are no theoretical calculations of this cross-section. Consequently, in this analysis, we adopt this cross-section as the cross-section for helium modified by the ratio of the cross-sections for charge exchanges of protons with neutral neon to this cross-section for hydrogen. While this is certainly a crude approximation, we believe it is sufficient for the purpose of this paper.

\begin{figure}
	\epsscale{.6}
	\plotone{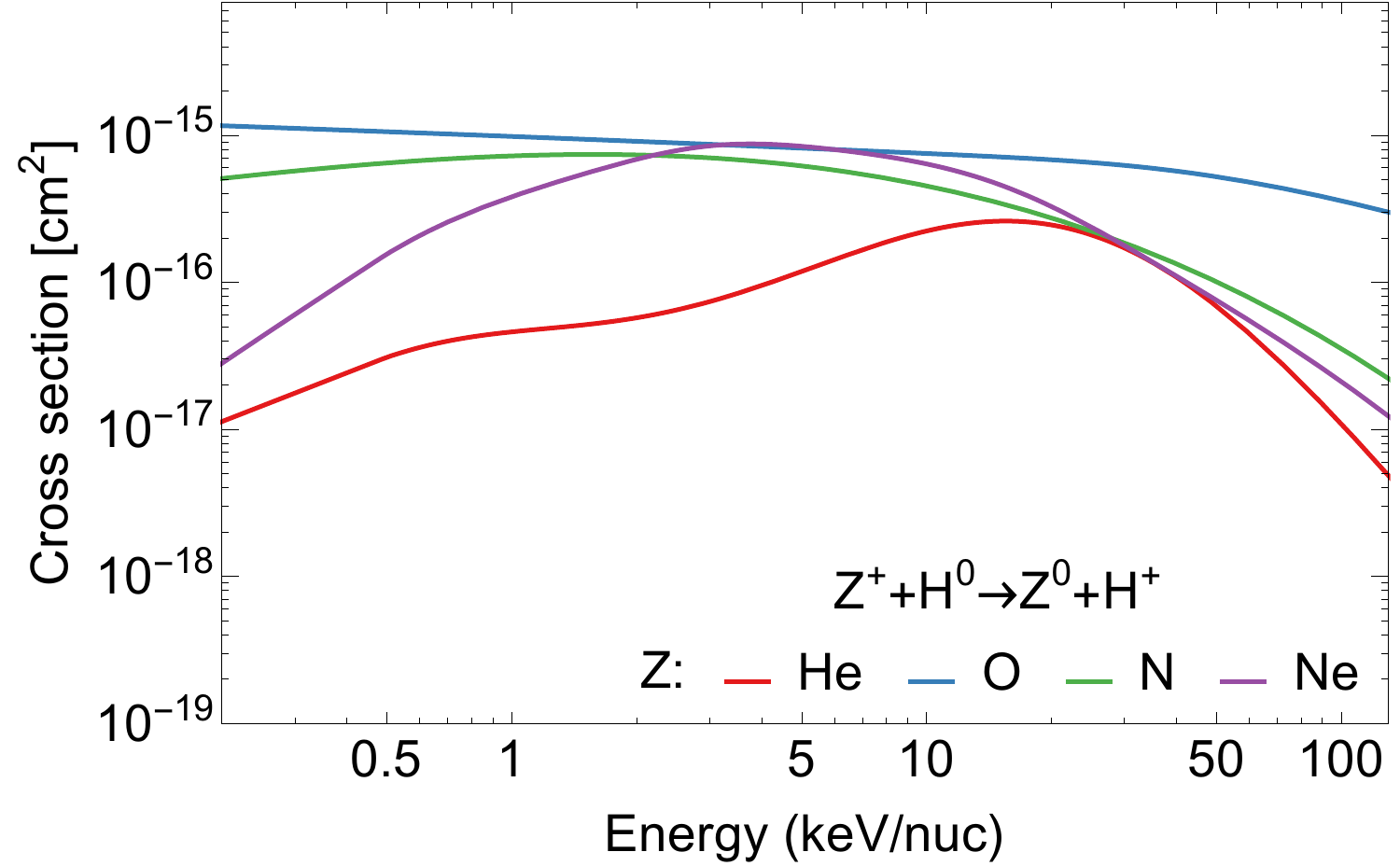}
	\caption{Cross sections for the charge exchanges between single ionized ions and hydrogen atoms as a function of projectile energy. Colors represent the considered chemical elements. \label{fig:cxin}}
\end{figure}

Figure~\ref{fig:cxin} shows the used cross-sections for neutralization of single ionized ions on neutral hydrogen atoms. The cross-section for oxygen is only slightly changing over the considered energy range. All cross sections, except for helium, have a similar magnitude for a few keV/nuc ions. 

\section{Ionization on carbon foils}\label{app:cfoils}
\begin{figure}
	\epsscale{.6}
	\plotone{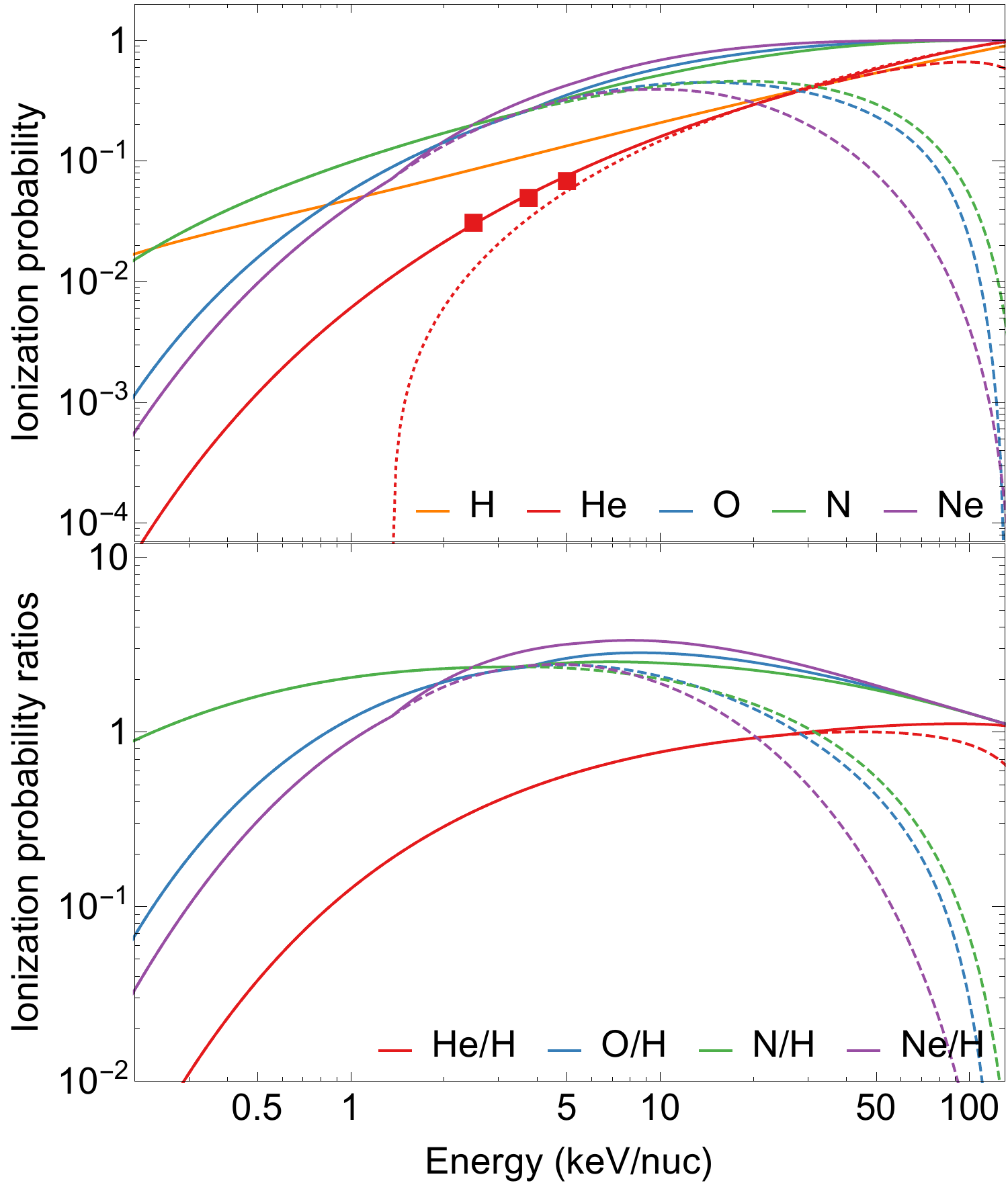}
	\caption{\emph{Upper panel:} ionization probabilities of atoms of various chemical elements passing through thin carbon foils as a function of atom energy. The dashed lines show the probability of single ionization, and the solid lines the probability of ionization to all possible ionization states. The dotted red line results from the original parameters given by \citet{gonin_1995} for helium, and the squares are measurements from \citet{allegrini_2014}. \emph{Lower panel:} ratios of the ionization probabilities for heavy atoms to the ionization probability of hydrogen. \label{fig:cfoils}}
\end{figure}

Probabilities of atom ionizations on carbon foils are functions of energy, different for each chemical elements. Here, the model by \citet{gonin_1994a, gonin_1994} is employed. The ionization probabilities for atoms passing through these foils resulting from this model are presented in Figure~\ref{fig:cfoils}. The figure shows (1) probabilities that atom escaping the foil is singly ionized (dashed lines) and (2) a sum over all ionized states (solid lines). For lower energies, the probability of multiple ionization is small, but it becomes important for energies $\gtrsim$10 keV/nuc for all species except for helium. 

The model derived by \citet{gonin_1994a, gonin_1994} is a semi-empirical model, in which parameters were adjusted to match experimental results that were available at this time. The ionization probability is small for low energy helium, and the model results do not agree with recent measurements by \citet{allegrini_2014}. This disagreement is probably a result of the lack of low energy measurements in the original derivation. Since this discrepancy may influence the results at the lowest energies, in this analysis, the parameters of the model were modified to reproduce more recent data. Namely, the values of parameters $k_\mathrm{P}^1/k_\mathrm{F}$ and $a_1$ are set as 0.99 and 0.92, respectively. These values allow for agreement with the recent measurements and simultaneously do not substantially change the values for higher energies.

Figure~\ref{fig:cfoils} also shows ratios of ionization probabilities for different elements in relation to those for hydrogen. For helium atoms, the probability is smaller than for hydrogen atoms for energies $\lesssim$30 keV/nuc, and the ratio is $\sim$0.1 for 1 keV/nuc, but it monotonically grows with energy and it is $\sim$0.8 for 10 keV/nuc and $\sim$1.1 for 100 keV/nuc. For other considered species, the probability is typically larger (up to $\sim$3 times) than for hydrogen for energies $\gtrsim$1 keV/nuc. 

\bibliographystyle{aasjournal}
\bibliography{library}


\end{document}